\documentclass[12pt]{article}
\pdfoutput=1 

\usepackage{cancel,slashed}
\usepackage{bbm}
\usepackage{amssymb}
\usepackage{amsfonts}
\usepackage{amsmath}
\usepackage{graphicx}
\usepackage{cite}

\usepackage{latexsym}
\usepackage{color}
\usepackage{xypic}
\usepackage{cancel,slashed}
\usepackage[linktocpage]{hyperref}
\usepackage{color}
\usepackage{transparent}
\usepackage{footmisc}

\makeatletter
\def\@xfootnote[#1]{%
  \protected@xdef\@thefnmark{#1}%
  \@footnotemark\@footnotetext}
\makeatother

\newcommand{\bmat}{\left(\begin{array}}
\newcommand{\emat}{\end{array}\right)}

\def\p{\partial}
\def\a{\alpha}

\def\g{\gamma}

\def\om{\omega}
\def\Om{\Omega}

\def\-{\hphantom{-}}

\def\s2{\frac{1}{\sqrt2}}

\def\oh{\frac{1}{2}}
\def\beq{\begin{equation}}
\def\eeq{\end{equation}}
\def\beqa{\begin{eqnarray}}
\def\eeqa{\end{eqnarray}}
\def\D{{\rm D}}
\def\im{{\rm Im \,}}
\def\re{{\rm Re \,}}
\def\tr{{\rm tr \,}}

\def\Dsl{\,\raise.15ex\hbox{/}\mkern-13.5mu D} 

\def\Db{{\bar D}}

\def\CK {{\cal K}}

\def\CO {{\cal O}}

\def\re{\mbox{Re}}
\def\im{\mbox{Im}}
\def\tr{\mbox{Tr}}

\def\be{\begin{equation}}
\def\ee{\end{equation}}
\def\bea{\begin{eqnarray}}
\def\eea{\end{eqnarray}}
\def\raw{\rightarrow}
\def\Raw{\Rightarrow}

\def\IC{\mathbb{C}}

\def\IR{\mathbb{R}}

\def\oh{\frac{1}{2}}
\def\a{{\alpha}}

\def\eps{{\epsilon}}

\def\lam{{\lambda}}
\def\Om{{\Omega}}
\def\om{{\omega}}
\def\sig{{\sigma}}

\def\g{{\gamma}}

\def\p{{\partial}}

\def\w{{\wedge}}

\newcommand{\conj}{\overline}
\newcommand{\mr}{\mathrm}
\newcommand{\mc}{\mathcal}
\newcommand{\mf}{\mathbf}
\newcommand{\mb}{\mathbb}
\newcommand{\mg}{\mathfrak}
\newcommand{\pb}{\conj{\p}}
\newcommand{\xb}{{\conj{x}}}
\newcommand{\yb}{{\conj{y}}}
\newcommand{\zb}{{\conj{z}}}

\newcommand{\phib}{{\conj{\phi}}}
\newcommand{\Ab}{{\conj{A}}}

\def\sm2{{\mbox{\small 2}}}

\begin{document}
\pagestyle{plain}

\makeatletter
\@addtoreset{equation}{section}
\makeatother
\renewcommand{\theequation}{\thesection.\arabic{equation}}
\pagestyle{empty}
\rightline{IFT-UAM/CSIC-16-082}
\vspace{0.5cm}
\begin{center}
\Huge{{T-branes and $\alpha$'-corrections}
\\[10mm]}
\large{Fernando Marchesano and Sebastian Schwieger \\[10mm]}
\small{
 Instituto de F\'{\i}sica Te\'orica UAM-CSIC, Cantoblanco, 28049 Madrid, Spain 
\\[15mm]} 
\small{\bf Abstract} \\[10mm]
\end{center}
\begin{center}
\begin{minipage}[h]{15.0cm} 

We study $\alpha$'-corrections in multiple D7-brane configurations with non-commuting profiles for their transverse position fields. We focus on T-brane systems, crucial in F-theory GUT model building. There $\a'$-corrections modify the D-term piece of the BPS equations which, already at leading order, require a non-primitive Abelian worldvolume flux background. We find that $\a'$-corrections may either {\it i)} leave this flux background invariant, {\it ii)} modify the Abelian non-primitive flux profile, or {\it iii)} deform it to a non-Abelian profile. The last case typically occurs when primitive fluxes, a necessary ingredient to build 4d chiral models, are added to the system. We illustrate these three cases by solving the $\a'$-corrected D-term equations 
in explicit examples, and describe their appearance 
in more general T-brane backgrounds. Finally, we discuss implications of our findings for F-theory GUT local models.

\end{minipage}
\end{center}
\newpage
\setcounter{page}{1}
\pagestyle{plain}
\renewcommand{\thefootnote}{\arabic{footnote}}
\setcounter{footnote}{0}


\tableofcontents


\section{Introduction}
\label{s:intro}

One important property of D-branes is that they greatly enhance the possibilities to build different 4d string theory vacua and, when present, they dictate to large extent the phenomenological interest of such vacua \cite{thebook}. 

Pictorially, we are used to think of D-branes as dynamic objects wrapped on submanifolds of a certain compactification manifold. While this intuition may be accurate for a single, isolated brane, systems of multiple D-branes are known to be richer than a simple sum of submanifolds, allowing for new configurations that can be understood in terms of non-Abelian bound states \cite{Myers:2003bw}. In this respect, attention has been given lately to configurations which can be thought of as a non-Abelian deformation of coincident or intersecting D-brane systems \cite{Donagi:2003hh,Hayashi:2009bt,Cecotti:2010bp}, dubbed T-branes in \cite{Cecotti:2010bp}. Such T-brane configurations are particularly interesting when they refer to systems of 7-branes, to which most of the subsequent formal developments are related \cite{Donagi:2011jy,Donagi:2011dv,Marsano:2012bf,Anderson:2013rka,Collinucci:2014qfa,Collinucci:2014taa,Collinucci:2016hpz,Bena:2016oqr}. 

Indeed, as follows from the scheme introduced in \cite{Donagi:2008ca,Beasley:2008dc,Beasley:2008kw,Donagi:2008kj}, in F-theory GUT models Yukawa couplings are computed in terms of a 7-brane super Yang-Mills theory with a non-Abelian group $G$, a Higgs/transverse-position field $\Phi$ and a gauge vector $A$. By specifying the internal profiles for $\Phi$ and $A$ around certain local patches, one is able to compute the physical Yukawas of each model, see \cite{Heckman:2008qa,Hayashi:2009ge,Randall:2009dw,Font:2009gq,Cecotti:2009zf,Conlon:2009qq,Marchesano:2009rz} for details. In this approach, in order to naturally achieve up-type quark masses with one family much heavier than the others one is forced to consider $G = E_{n}$ and a T-brane profile for $\Phi$ \cite{Cecotti:2010bp}. 
 Such general setup can been implemented in different local models in order to achieve realistic Yukawa couplings, see \cite{Chiou:2011js,Font:2013ida,Marchesano:2015dfa,Carta:2015eoh}.

It is nevertheless important to notice that the 7-brane SYM theory described in \cite{Donagi:2008ca,Beasley:2008dc,Beasley:2008kw,Donagi:2008kj} is subject to $\a'$-corrections. In the case of multiple D7-branes such $\a'$-corrections are encoded in the non-Abelian DBI+CS actions, and their effect can in principle be extracted directly from there. In practice it is however simpler to see how these correction modify the BPS equations for multiple D7-branes, and then analyse the configurations that solve the corrected equations. The purpose of this paper is to apply this strategy to analyse $\a'$-corrections in T-brane systems of D7-branes, including all those ingredients that appear in F-theory GUT model building.

Since D7-branes wrapping holomorphic four-cycles are examples of B-branes, we expect that $\a'$-corrections do not modify their F-term equations and only affect their D-term BPS equations. In other words, if we describe the corrected BPS equations as a Hitchin system, the holomorphic 7-brane data will remain unaffected and $\a'$-corrections will only modify the stability condition \cite{Minasian:2001na}. This result, which we review from the viewpoint of \cite{Butti:2007aq,Marchesano:2010bs}, allows to solve for $\a'$-corrected T-brane backgrounds with the same strategy used in \cite{Cecotti:2010bp}: we first define their holomorphic data and then solve the D-term equation in terms of a complexified gauge transformation acting on $\Phi$ and $A$. We will then see that $\a'$-corrections will not only change the initial T-brane profile quantitatively, but also qualitatively. 

Indeed, a standard class of T-brane configurations features a Higgs field $\Phi$ along a set of non-commuting generators $E_i$ and a non-primitive worldvolume flux of the form
\be
F \, =\, -i \p \pb f \, P
\label{npflux}
\ee
that solves the classical D-term equation. Here $P$ is a Cartan generator of the gauge group $G$, while $f$ is a function of the 7-brane coordinates that solves a certain differential equation and that also enters in the profile for $\Phi$ \cite{Cecotti:2010bp}. While non-trivial, this Abelian profile for $F$ is relatively simple, in the sense that it could involve several, non-commuting generators of $G$. In this paper we will consider the $\a'$-corrected version of this class of systems. As a general result we find that several things can happen:

\begin{itemize}

\item[{\it i)}]
In the most simple example of this setup, which preserves eight supercharges, the same background is also a solution of the $\a'$-corrected D-term equations. 

\item[{\it ii)}]
We may lower the amount of supersymmetry to four supercharges by 

\begin{itemize}

\item[{\it a)}] modifying the Higgs background as $\Phi \raw \Phi + \Delta \Phi$, with $[\Phi, \Delta \Phi] = [F,  \Delta \Phi] = 0$, 

\item[{\it b)}] introducing a primitive worldvolume flux $H$ that commutes with $\Phi$ and $F$.

\end{itemize}

Ignoring $\a'$-corrections {\it a)} and {\it b)} do not modify the T-brane piece of the background. However, taking $\a'$-corrections into account the profile for the function $f$ is modified.

\item[{\it iii)}]
If we perform {\it a)} and {\it b)} simultaneously while preserving four supercharges then, in general, (\ref{npflux}) may not solve the $\a'$-corrected D-term equations and the non-primitive flux $F$ will have to develop new components along the non-Cartan generators $E_i$. The T-brane profile for $\Phi$ will also become more involved. 

\end{itemize}

Interestingly, {\it a)} and {\it b)} are standard features that one needs to implement in local F-theory GUTs in order to engineer realistic 4d chiral models \cite{Font:2013ida,Marchesano:2015dfa,Carta:2015eoh}. One may therefore expect that, in general, the description of T-brane systems leading to realistic F-theory models will be qualitatively modified when taking into account the effect of $\a'$-corrections, at least at the level of non-holomorphic data. 

The paper is organised as follows. In section \ref{s:Dterms} we derive how $\alpha'$-corrections enter systems of multiple D7-branes, and in particular how they modify their D-term equations. In section \ref{s:inter} we solve such $\a'$-corrected D-term equations for system of intersecting D7-branes, relating the corrections to the pull-back on each individual D7-brane embedding. Then, in section \ref{s:Tbrane}, we turn to solve the $\a'$-corrected D-term equations for simple T-brane backgrounds, which already illustrate the three cases described above. In section \ref{s:gen} we discuss how to solve $\a'$-corrected D-term equations in more general T-brane systems and how the same phenomena arise in there. In section \ref{s:app} we briefly comment on the implications of our finding for some local F-theory GUT models. We draw our conclusions in section \ref{s:conclu}.

Several technical details have been relegated to the Appendices. Appendix \ref{ap:CS} contains an alternative derivation of the $\a'$-corrected D-term equations by means of the non-Abelian Chern-Simons action. Appendix \ref{ap:toda} shows that $\a'$-corrections are trivial for certain T-brane systems with globally nilpotent Higgs field. 
Appendix \ref{ap:genAnsatz} shows how adding non-Cartan flux backgrounds can solve the corrected D-term equations in the T-brane backgrounds of section \ref{s:Tbrane} that correspond to case {\it iii)}, at least to next-to-leading order in the $\a'$-expansion. Appendix \ref{ap:more} shows the results of the analysis of section \ref{s:Tbrane} applied to further $SU(2)$ T-brane backgrounds.

\section{D7-branes, D-terms and their $\alpha$'-corrections}
\label{s:Dterms}

Let us consider type IIB string theory compactified on a Calabi-Yau threefold $X_3$, and then quotiented by an orientifold action such that the presence of O3/O7-planes is induced. In order to cancel the related RR charge of these orientifold content one may add different stacks of D3-branes and D7-branes, the latter wrapping four-cycles $\mc{S}_a \subset X_3$ in the internal space and with internal worldvolume fluxes $F$ switched on along $\mc{S}_a$. 

In the simplest configuration that one may consider each stack would only involve a single D7-brane, wrapping a collection of different, isolated four-cycles $\{\mc{S}_a\}$. For each of these D7-branes one can check if the energy is minimised by looking at its BPS conditions, which amount to require that the four-cycle $\mc{S}$ is holomorphic that the worldvolume flux threading it is a primitive (1,1)-form in $\mc{S}$ \cite{Marino:1999af,Gomis:2005wc,Martucci:2005ht}.\footnote{In our conventions $\mc{S}$ is calibrated by $-J^2$ and so a BPS worldvolume flux is self-dual $F = *_{\mc{S}} F$.} These BPS conditions are captured by the following functionals \cite{Jockers:2005zy,Martucci:2006ij}
\begin{align}
\label{Wab}
	W &= \int_{\Sigma_5}  \mr{P} \left[\Omega_0 \w e^{-B} \right] \w e^{\lambda F} \\
	D &= \int_\mc{S} \mr{P} \left[ \mr{Im}\, e^{iJ} \w e^{-B} \right] \w e^{\lambda F}
\label{Dab}
\end{align}
that in 4d are respectively interpreted as a superpotential and D-term for each D7-brane.  Here $J$ is the K\"ahler form and $\Omega_0 = e^{\phi/2} \Omega$ a holomorphic (3,0)-form  in $X_3$, normalised such that $\frac{1}{6} J^3 = -\frac{i}{8} \Omega\wedge \bar{\Omega}$. In addition, $B$ is the internal B-field, $F=dA$ the worldvolume flux and $\lambda = 2\pi \alpha'$. Finally, $\Sigma_5$ is a five-chain describing the deformations of the four-cycle $\mc{S}$, which infinitesimally can also be parameterised by the complex position coordinates $\Phi^i$, and $P[\dots ]$ stands for the pull-back on the D7-brane worldvolume, namely
\begin{align}
\mr{P} \left[ V_\mu \mr{d} z^\mu \right]_\alpha = V_\alpha + \lambda V_i \p_\alpha \Phi^i
\end{align}
with $\alpha$ a coordinate in $\mc{S}$. 

More generally, one would consider configurations involving stacks of several 7-branes, with non-Abelian bundles on them and wrapping four-cycles that intersect each other. On a given patch of the internal manifold one can describe such configurations in terms of a  8d twisted super Yang-Mills theory with a given non-Abelian symmetry group $G$ \cite{Donagi:2008ca,Beasley:2008dc,Beasley:2008kw,Donagi:2008kj}. The bosonic field content of this theory is given by a gauge field $A$ and a Higgs-field $\Phi$ transforming in the adjoint of $G$, and whose background profiles will break $G$ to a smaller gauge symmetry group. In this paper we are interested in configurations in which the profile for $\Phi$ is intrinsically non-Abelian, and more precisely in the kind of profiles considered in \cite{Hayashi:2009bt,Cecotti:2010bp,Donagi:2011jy,Anderson:2013rka,Collinucci:2014qfa,Collinucci:2014taa} and dubbed T-branes in \cite{Cecotti:2010bp}. 

Just like in the Abelian case, the non-Abelian profiles for $\Phi$ and $A$ need to satisfy certain equations of motion that are captured by 7-brane functionals. In order to describe the non-Abelian generalisation of (\ref{Wab}) and (\ref{Dab}) one may proceed as follows \cite{Butti:2007aq,Marchesano:2010bs}.\footnote{See \cite{Minasian:2001na}  for a previous, alternative derivation of these equations.} First one uses the equations of motion of the background to locally write $\Omega_0 \wedge e^B = d\gamma$, and so rewrite the integral in (\ref{Wab}) as $\int_{\mc{S}} P[\gamma] \wedge e^{\lam F}$. Then one observes that, since both $W$ and $D$ have both the  form of the D7-brane Chern-Simons action, their non-Abelian generalisation should go along the same lines as described in \cite{Myers:1999ps}. More specifically, we replace the derivatives in the pull-back by gauge-covariant ones and symmetrise over the gauge indices. We finally obtain
\begin{align}
	W &= \int_\mc{S} \mr{STr} \Big\{ \mr{P} \left[ e^{i \lambda \iota_\Phi \iota_\Phi} \gamma \right] \w e^{\lambda F} \Big\} \label{s:Dterms/Wpurespin}\\
	D &= \int_\mc{S} \mr{S} \Big\{ \mr{P} \left[ e^{i \lambda \iota_\Phi \iota_\Phi } \mr{Im}\, e^{iJ}  \w e^{-B} \right] \w e^{\lambda F} \Big\}. \label{s:Dterms/Dpurespin}
\end{align}
where $\iota_\Phi$ stands for the inclusion of the complex Higgs field $\Phi$, and $\mr{S}$ for symmetrisation over gauge indices. Just like eqs.(\ref{Wab}) and (\ref{Dab}), these functionals describe the D-brane BPS equations whenever the approximations leading to the D-brane DBI + CS actions hold, namely internal volumes with are large and slowly varying profiles for $\Phi$ and $F$ in string length units. In this regime the D-term functional (\ref{s:Dterms/Dpurespin}) should take into account all the $\a'$-corrections to the BPS equations for a non-Abelian system of D7-branes.\footnote{That is, if we neglect higher derivative corrections of the Riemann tensor. After taking such curvature corrections into account one expects a non-Abelian D-term of the form \cite{Minasian:2001na}
\be \nonumber
D = \int_{\mc{S}} \mr{P} \left[ \mr{Im}\, e^{iJ} \w e^{-B} \right] \w e^{\lambda F'} \w \sqrt{\hat{\mc{A}}(\mc{T})/\hat{\mc{A}}(\mc{N})} 
\ee 
with $\hat{\mc{A}}$ the A-roof genus of the tangent $\mc{T}$ and normal $\mc{N}$ bundles, and $F' = F - \oh F_{\mc{N}}$ with $F_{\mc{N}}$ the normal bundle curvature \cite{Green:1996dd,Cheung:1997az,Minasian:1997mm,Morales:1998ux,Haack:2006cy}. Here $\sqrt{\hat{\mc{A}}(\mc{T})/\hat{\mc{A}}(\mc{N})} = 1 - \frac{1}{48}[ p_1(\mc{T}) -  p_1(\mc{N})] + \dots$ with $p_1$ the real four-form given by the first Pontryagin class. Note that this correction does not affect the Abelian D-term but it is non-trivial in the non-Abelian case. In the following we will consider a local patch in which the K\"ahler metric is locally flat, and therefore take $p_1=0$ and $F' = F$. It would be interesting to see if our results could change qualitatively when these curvature corrections become important.}

In order to bring these expressions to a more familiar form let us introduce local complex coordinates $x,y,z$ and take the four-cycle $\mc{S}$ along the locus $\{z=0\}$ -- that is $x$ and $y$ are the coordinates of $\mc{S}$. In this local description the Higgs field is given by
\be
\Phi \equiv \phi \frac{\p}{\p z} + \phib \frac{\p}{\p \conj{z}}.
\ee
where $\phi$ is a matrix in the complexified adjoint representation of $G$ and $\phib$ its Hermitian conjugate. Locally we may also take $\gamma \equiv z\, \mr{d} x \w \mr{d} y$, such that  in particular we have $\iota_\Phi \gamma = 0$. Performing a normal coordiante expansion of $\g$ and plugging it into \eqref{s:Dterms/Wpurespin} then gives
\begin{align}
	W &= \lambda^2 \int_{\mc{S}} \mr{Tr} \left\{ \phi\, dx\w dy \w F \right\}\, =\, \lambda^2 \int_{\mc{S}} \mr{Tr} \left\{ \iota_{\Phi} \Omega \w F \right\}.
\end{align}
which is the 7-brane superpotential considered in \cite{Donagi:2008ca,Beasley:2008dc,Beasley:2008kw}.\footnote{Notice that in these references the two-form $ \iota_{\Phi} \Omega$ is denoted by $\Phi$.} Crucially, the integrand does not depend on $\lambda$, which implies that the F-term conditions are entirely topological and receive no $\alpha'$-corrections. 

We will now see that this is not the case for the D-terms (\ref{s:Dterms/Dpurespin}), which are evaluated as
\begin{align}
	D = \int_{\mc{S}} \mr{S} &\Bigg\{ \lambda P[J] \w F - \frac{i \lambda}{6} \iota_\Phi \iota_\Phi J^3  + \frac{i \lambda^3}{2} \iota_\Phi \iota_\Phi J \w F \w F \nonumber\\
	&- \mr{P}[J \w B] - i \lambda^2 \iota_\Phi \iota_\Phi ( J \w B ) \w F + \frac{i \lambda}{2} \iota_\Phi \iota_\Phi (J \w B^2)
	\Bigg\},
\end{align}
where we have kept terms of all orders in $\lambda$ in this expansion.\footnote{Including curvature corrections there would be an extra term of the form $\frac{i\lam}{48} \iota_\Phi\iota_\Phi J [ p_1(\mc{T}) -  p_1(\mc{N})]$.} In our local patch we may take the flat space K\"ahler form to be
\begin{align}
	J = \underbrace{\frac{i}{2} \mr{d}x \w \mr{d} \xb + \frac{i}{2} \mr{d}y \w \mr{d} \yb}_{=: \omega} + 2i \mr{d}z \w \mr{d} \conj{z},
\label{Jlocal}
\end{align}
decompose the background B-field as  $B \equiv B \big|_\mc{S} + B_{z\zb} \mr{d}z \w \mr{d}\zb$ and write $\mc{F} = \lambda F -B\big|_\mc{S} $, yielding
\begin{align}
	D = \int_{\mc{S}} \mr{S} &\Bigg\{ \mr{P}[J] \w \mc{F} + \frac{i \lambda}{2} \left( \iota_\Phi \iota_\Phi J \right) \left( \mc{F}^2 - \omega^2 \right) 
	- i \lambda \left( \iota_\Phi \iota_\Phi B \right) \omega \w \mc{F} - \omega \w \mr{P} [ B_{z\zb} \mr{d}z \w \mr{d}\zb ] \nonumber
	\Bigg\}.
\end{align}
Here we defined the Abelian pull-back $\omega$ to $\mc{S}_4$ as indicated in (\ref{Jlocal}), such that we have
\begin{align}
	\iota_\Phi \iota_\Phi J &= 2i [\phi, \phib] \nonumber\\
	\iota_\Phi \iota_\Phi J^3 &= 6i [\phi, \phib] \omega^2. \nonumber
\end{align}
To proceed we note that $2i [\phi, \phib]$ is a zero-form and secondly, that $6i [\phi, \phib] \omega^2$ has no transverse legs to $\mc{S}$. That is, in both cases the pull-back P acts trivially. Lastly, one may compute
\begin{align}
	\mr{P}[J] = \omega + 2i \lambda^2 (\D \phi) \w (\Db \phib).
\end{align}
so at the end we have that the D-term equations amount to ${D} =0$ with 
\begin{align}
	D = \int_{\mc{S}} \mr{S} &\Bigg\{ \omega \w \mc{F} + \lambda^2 \D \phi \w \conj{\D} \phib \w \left( 2i \mc{F} - B_{z\zb} \omega \right) + \lambda \left[ \phi, \phib \right] \left(\omega^2 - \mc{F}^2 -  i B_{z\zb} \omega \w \mc{F}\right)  \label{s:Dterms/D-termB}
	\Bigg\}.
\end{align}
For vanishing $B$-field, this simplifies to
\begin{align}
	D = \lambda \int_{\mc{S}} \mr{S} &\Bigg\{ \omega \w F + 2i \lambda^2 \D \phi \w \conj{\D} \phib \w F + \left[ \phi, \phib \right] \left( \omega^2 - \lambda^2 F^2 \right) 
	\Bigg\}. \label{s:Dterms/D-term}
\end{align}
These expressions reproduce those found in \cite{Minasian:2001na}, and can be recovered by analysing the non-Abelian Chern-Simons action of a stack of D7-branes, as discussed in Appendix \ref{ap:CS}.  

Note that both terms at leading order in $\lambda$, namely $ \omega \w F +  \left[ \phi, \phib \right] \omega^2$, are purely algebra valued. Crucially, this is not the case anymore when we include higher orders, because these additional terms contain products of generators. From the original formula in \eqref{Dab} it is clear that these products have to be understood in the same way as in the exponentiation map, which implies that for matrix algebras $\mg{g} \subset GL({n,\mb{C}})$ they are simply the matrix products in the fundamental representation of said algebra. Taking into account the symmetrisation procedure, we end up considering terms of the form
\begin{align}
	\mr{S} \big\{ T_1 \dots T_n \big\} = \frac{1}{\text{\# of perm.}} \sum_{\text{all perm. }\sigma} T_{\sigma_1} \cdots T_{\sigma_n}.
\end{align}
Formally speaking, including higher order corrections in $\lambda$ means that the D-terms are valued in the universal enveloping algebra $U(\mg{g})$ rather than $\mg{g}$ itself.

\section{$\alpha$'-corrections for intersecting branes}
\label{s:inter}

To get some intuition on the meaning of the $\a$' corrections on D-terms, let us first consider the case where the Higgs field $\phi$ and the gauge flux $F$ can be diagonalised, as is for the case of intersecting D7-brane backgrounds. Then the D-term equations amount to
\begin{align}
	D &= \lambda \int_{\mc{S}} \mr{P}_{\rm ab}[J] \w F \,= \, \lambda \int_{\mc{S}} \left( \omega + 2 i \lambda^2 \p \phi \w \pb \phib \right) \w F,
\end{align}
that is to say the $\alpha'$-corrections are given entirely by the abelian pull-back of the K\"ahler-form $J$ to $\mc{S}$, $\mr{P}_\mr{ab}[J] \equiv \left( \omega + 2 i \lambda^2 \p \phi \w \pb \phib \right)$. This implies that flux needs to be primitive with respect to this pull-back rather than with respect to $\omega \equiv J|_{\mc{S}} = \frac{i}{2} \left( \mr{d}x \w \mr{d}\xb + \mr{d}y \w \mr{d}\yb \right)$, the difference being the $\a^\prime$ corrections to the D-term. 

Let us be more specific and consider the background
\begin{align}
	\phi= \left(\begin{matrix} \mu^2 x & 0\\ 0 & -\mu^2 x \end{matrix}\right)
\end{align}
and a flux $F$ that commutes with $\phi$. Namely we have
\be
F\, =\, F_{x\bar{x}}\, dx \wedge d\bar{x} + F_{y\bar{y}}\, dy \wedge d\bar{y} + F_{x\bar{y}}\, dx \wedge d\bar{y} + F_{y\bar{x}}\, dy \wedge d\bar{x}
\ee
where $F = F^\dagger$ imposes $F_{x\bar{y}} = F_{y\bar{x}}$ and a reality condition for $F_{x\bar{x}}$, $F_{y\bar{y}}$. In particular, due to our Ansatz these components must be of the form $i (a \sig_3 + b \mf{1})$, with $a$, $b$ real functions. 

Imposing that $dF=0$ and the leading order D-term condition $\om \wedge F = 0$ sets these functions to be constant and such that $F_{x\xb} = -F_{y\yb}$, while $F_{x\bar{y}}$ is constant but otherwise unconstrained. The latter is also true for the $\a'$-corrected D-term constraint, while the relation between $F_{x\xb}$ and $F_{y\yb}$ is modified to
\begin{align}
	F_{x\xb} = -(1+4\lambda^2 |\mu|^4) F_{y\yb}, \label{s:inter/Dterm}
\end{align}
Notice that this condition reduces to the naive primitivity condition $F_{x\xb} + F_{y\yb} = 0$ in the limit $\lam \raw 0$, while for finite $\lam$ it gives a correction that grows with the complex parameter $\mu \in \IC$, $[\mu] = L^{-1}$.

Physically, the $\a'$-corrected D-term condition is quite easy to understand. Indeed, notice that the Higgs-field vev in (\ref{s:inter/Dterm}) describes an $SU(2)$ gauge theory which is broken completely over generic loci, and in particular there is no D7-brane on the naive gauge theory locus $\{z=0\}$. Instead we may compute the D7-brane loci via the discriminant $\det \left( z \cdot \mf{1} - \lambda \cdot \phi \right) = (z - \lambda \mu^2 x) (z + \lambda \mu^2 x)$, which indicates that the system contains two D7-branes located at $\{z=\pm\lambda \mu^2 x\}$ and $\mu^2$ is their intersection slope. A more suitable description can be obtained by passing to a new system of coordinates
\begin{align}
	u &\equiv z + \lambda \mu^2 x \\
	v &\equiv z - \lambda \mu^2 x \\
	w &\equiv y,
\end{align}
in which the branes loci are given by $\{u=0\}$ and $\{v=0\}$, and then analysing each of the D7-branes individually in term of their Abelian D-terms. For instance, to have primitive flux along the D7-brane located at $\{u=0\}$ translates into
\begin{align}
	0 &=  J|_{\{u=0\}} \w F \\
	&\Rightarrow F_{v\bar{v}} = -\left( \frac{1}{4\lambda^2 |\mu|^4} + 1 \right) F_{w\bar{w}} \\
	&\Rightarrow F_{x\xb} = -\left(1 + 4 \lambda^2 |\mu|^4 \right) F_{y\yb}
\end{align}
and similarly for $\{v=0\}$. This is precisely the result we obtained earlier in \eqref{s:inter/Dterm} from the perspective of the gauge theory on $\{z=0\}$. So intuitively the D-term equations in this description just tells us that the flux should be primitive along the actual brane world-volumes, rather than the locus $\mc{S}$ from which we describe the parent gauge theory.

\section{$\alpha$'-corrections in simple T-brane backgrounds}
\label{s:Tbrane}

After seeing the effect of $\a'$ corrections for intersecting D7-branes, let us investigate which types of effects we receive for T-brane backgrounds. In general, these backgrounds are such that $[\phi, \bar{\phi}] \neq 0$ and so a non-primitive flux $F$, satisfying $F\wedge \omega \neq 0$, is needed to solve the D-term equations at leading order \cite{Cecotti:2010bp}. 

In order to find BPS solutions for these backgrounds one may apply the strategy outlined in \cite{Cecotti:2010bp}. Namely, one first defines the T-brane Higgs background in a unphysical holomorphic gauge \cite{Font:2009gq,Cecotti:2009zf}
\be
A^{(0,1)} = 0 \qquad \qquad \bar{\p} \phi_\mr{hol} = 0 
\ee
and then rotate these fields by a complexified gauge transformation of the symmetry group $G$ 
\be
A^{(0,1)} \, \raw \, A^{(0,1)} + ig\bar{\p} g^{-1}\qquad \quad \phi \, \raw \, g \phi g^{-1}
\label{gaugetr}
\ee
in order to attain a unitary gauge in which the D-term condition is satisfied. In the following we will apply this same strategy to solve for the $\a'$-corrected D-term equations. We will consider two simple examples in which the leading order non-primitive flux lies in the Cartan subalgebra of the symmetry group $G$, as this also simplifies the Ansatz to solve the D-term equations at higher order in $\a'$.

\subsection{A simple SU(2) background}
\label{s:Tbrane/SU2}

Let us first analyse a simple $SU(2)$-background already considered in \cite{Cecotti:2010bp} where the Higgs field profile in the holomorphic gauge reads
\begin{align}
\phi_\mr{hol} = m \left(\begin{matrix} 0 & 1 \\ a x & 0 \end{matrix}\right)\, =\, -imE^++ im axE^- \label{su2hol}
\end{align}
where $m, a \in \IC$ and $[m] = [a] = L^{-1}$, and the generators $E^\pm$ are defined in Appendix \ref{ap:genAnsatz}. This time the discriminant gives the D7-brane locus $z^2 = \lam^2 a m^2 x$. Moreover, since we have $\det \phi_\mr{hol} = -m^2 a x$, we see that this is a reconstructible brane background according to the definition given in \cite{Cecotti:2010bp}. To solve the D-term equations we proceed as above and pass to a unitarity gauge via a complexified gauge transformation in SU(2). More precisely we take
\be
g\, =\, e^{\frac{f}{2} \sig_3}
\label{gf}
\ee
which implies that in the unitarity gauge the D7-brane backgrounds reads
\begin{align}
\phi &= m \left(\begin{matrix} 0 & e^f\\ a x e^{-f} & 0 \end{matrix}\right)\, , \label{u2back}\\
F &= -i \p \pb f \cdot \sigma_3  \, .
\end{align}
At leading order in $\lam$ the D-term equations read
\be
\left( \p_x \pb_\xb + \p_y \pb_\yb \right) f\, \sig_3 = [\phi, \bar{\phi}] \quad \Raw \quad \left( \p_x \pb_\xb + \p_y \pb_\yb \right) f =  |m|^2 \left( e^{2f} - |a x|^2 e^{-2f} \right).
\label{painleIII}
\ee
Finding $f$ at this level amounts to solve a partial differential equation of Painlev\'e III type on the radial coordinate $|x|$, as has been already discussed in \cite{Cecotti:2010bp}. More precisely, we may solve it by making the Ansatz $f=f(|x|)$ and parametrise $x \equiv r e^{i \theta}$, yielding
\begin{align}
	\left( \frac{\mr{d}^2}{\mr{d}r^2} + \frac{1}{r} \frac{\mr{d}}{\mr{d}r}\right) f = |m|^2 \left( e^{2f} - |a|^2 r^2 e^{-2f} \right).
\end{align}
Redefining $e^{2f(r)} \equiv r |a| e^{2j(r)}$ further simplifies this to
\begin{align}
	\left( \frac{\mr{d}^2}{\mr{d}r^2} + \frac{1}{r} \frac{\mr{d}}{\mr{d}r}\right) j = |a| |m|^2 r \sinh (2j).
\end{align}
Finally we define $s \equiv \frac{2}{3} \sqrt{2 |a| |m|^2 r^3}$ such that we are left with
\begin{align}
	\left( \frac{\mr{d}^2}{\mr{d}s^2} + \frac{1}{s} \frac{\mr{d}}{\mr{d}s}\right) j = \sinh (2j),
\end{align}
which is the standard expression for a particular kind of Painlev\'e III equation analysed in \cite{McCoy:1976cd}. Finally, we may directly solve (\ref{painleIII}) asymptotically near $|x|=0$ by
\begin{align}
f = f_0 (x,\xb) &= \log c + c^2 |mx|^2 + \frac{|m|^2|x|^4}{4 c^2} \left( 2 |m|^2 c^6 - |a|^2 \right) + \dots 
\label{solf0}
\end{align}
with $c$ an arbitrary dimensionless parameter whose value should be close to $0.73$ if we want to avoid poles for large values of $|x|^2$ \cite{Font:2013ida}.

Let us now consider the $\a'$-corrected D-term equation. Applying (\ref{s:Dterms/D-term}) to this setup we obtain  the following equation
\begin{align}
\left( \p_x \pb_\xb + \p_y \pb_\yb \right) f & =  |m|^2 \left( e^{2f} - |ax|^2 e^{-2f} \right) \left(1 + 4 \lam^2 Q_f \right) + \lam^2 R [f,f]\, ,
\end{align}	
where
\bea
Q_f & = & (\p_x \pb_\xb f)(\p_y \pb_\yb f) - (\p_x \pb_\yb f)(\p_y \pb_\xb f) \label{Rfg} \\
R [f,g] & = & |m|^2 \left[ \left( 4 \p f \w \pb f e^{2f} + |a|^2 e^{-2f} (\p x - 2x \p f) \w ( \pb \xb - 2\xb \pb f) \right) \w \p\pb g \right]_{x\xb y\yb} \nonumber
\eea
describe the new operators that appear due to the $\a'$-corrections. Notice however that by keeping the Ansatz $f \equiv f (x,\bar{x})$ both $Q_f$ and $R[f,f]$ vanish identically and we are back to eq.(\ref{painleIII}). Therefore, the solution to the corrected D-term still amounts to $f = f_0 (x,\bar{x})$ and the above T-brane background does not suffer any modification due to $\a'$-corrections. Notice that in this case the T-brane background preserves $1/4$ of the supercharges of flat space. 
 Further examples of T-brane systems preserving eight supercharges are analysed in Appendix \ref{ap:toda}, again obtaining the result that $\a'$-corrections do not modify the background.

The analysis becomes more interesting if we consider a more general flux background, with a new component which will lower the amount of preserved supersymmetry. As usual we may consider adding such fluxes along generators that commute with the T-brane background. For instance we may add a worldvolume flux along the identity generator of $\mathfrak{u} (2)$, which could arise either from the D7-brane itself or form the pull-back of a bulk B-field. We first consider the case where this flux is
\be
H_1 \, =\, \im \left(\kappa \, dx \wedge d\bar{y}\right)\, \mf{1}
\label{H1}
\ee
with $\kappa \in \IC$ and $[\kappa] = L^{-2}$  parameterising the local flux density. At leading order in $\a'$, the vanishing D-term condition would allow for an arbitrary $\kappa$ without modifying the T-brane background, as the above flux is primitive. Its $\a'$-corrected counterpart, however, has non-trivial components along the generators $\sig_3$ and $\mf{1}$, implying two independent D-term equations. Namely
\begin{align}
\left( \p_x \pb_\xb + \p_y \pb_\yb \right) f & = |m|^2 \left( e^{2f} - |ax|^2 e^{-2f} \right) \left(1 + 4 \lam^2 Q_f + \lam^2 |\kappa|^2 \right) + \lam^2 R[f,f] \\
0 &= \mr{Re} \Big( |a|^2 e^{-2 f} \kappa x \p_y f \left(2 \xb \pb_\xb f - 1 \right) + 2 e^{2 f} \kappa \p_y f \pb_\xb f \Big) \nonumber
\end{align}
with the second line corresponding to the D-term constraint along the identity generator. 
Such equation is automatically satisfied if we again impose the Ansatz $f \equiv f (x,\bar{x})$, while the first one becomes
\begin{align}
\left( \p_x \pb_\xb + \p_y \pb_\yb \right) f & =  |m|^2 \left( e^{2f} - |ax|^2 e^{-2f} \right) \left(1 + \lam^2 |\kappa|^2 \right).  \label{kappasol}
\end{align}	
Hence, we are back to eqs.(\ref{painleIII}) and (\ref{solf0}) with the replacement
\be
m \, \raw\, m' = m \sqrt{1 + \lam^2 |\kappa|^2}.
\label{mprime}
\ee

Finally, let us consider the case where the flux background on the identity is
\be
H\, =\, H_1 + H_2 - i \p\bar{\p} h \, \mf{1} 
\label{Hflux}
\ee
where $H_1$ is again given by (\ref{H1}), and $H_2$ is an different piece of primitive constant flux
\be
H_2 \, =\, \rho \,  i \left(dx \wedge d\bar{x} - dy \wedge d\bar{y} \right) \, \mf{1} 
\label{H2}
\ee
with $\rho \in \IR$ and $[\rho] = L^{-2}$. In addition, we consider $h \, \equiv \, h(x, \xb, y, \yb)$ to be an arbitrary function that we may expand around the origin as a polynomial, starting at quadratic order. In addition, we write the gauge transformation (\ref{gf}) as the following expansion
\be
f \, =\, f_{0} (x,\bar{x}) + \sum_{i=1}^\infty (\lam \rho)^{2i}\,  f_i(x,\xb,y,\yb) 
\label{expf}
\ee
with $f_{0} (x, \bar{x})$ the solution found for $\rho =0$, which near the origin behaves as (\ref{solf0}) with the replacement (\ref{mprime}). 

In this case solving the D-term equations becomes more challenging, but one may perform a perturbative expansion on the dimensionless parameter $\lam \rho$ and keep the terms up to $\CO((\lam\rho)^2)$ in order to simplify them. On the one hand, for the D-term constraint along the generator $\sig_3$ we find
\begin{align}
(\p_x \pb_\xb + \p_y \pb_\yb ) f\, \sig_3 &=  [\phi,\phib] \left( 1 + 4 \lambda ^2 Q_H \right) ,
\label{Hsig3}
\end{align}
where now 
\begin{align}
Q_H &= \left( \p_x \pb_\xb h - \rho \right) \left( \p_y \pb_\yb h + \rho \right) - \left( \p_x \pb_\yb h - \frac{i}{2} \kappa \right) \left( \p_y \pb_\xb h - \frac{i}{2} \conj{\kappa} \right) .
\label{QH}
\end{align}
On the other hand, for the constraint along the identity we have 
\begin{align}\nonumber
(\p_x \pb_\xb + \p_y \pb_\yb ) h &= \lambda^2 \left[4|m|^2 \left( e^{2f} - |a x|^2 e^{-2f} \right)\p_x \pb_\xb f \Big( \p_y \pb_\yb h+ \rho \Big) -2 R [f,h + \rho |y|^2] \right]
\end{align}
with $R$ defined as in (\ref{Rfg}). We find the following solutions for $h$ at lowest orders in $\lam \rho$ and near the origin
\begin{align}
h & = \lambda^2 \rho  |mx|^2 \left( |m'x|^2 \left(|a|^2 + 2 c^6 |m'|^2 \right) -  \frac{2}{c^2} \left( |a|^2 - 2 c^6 |m'|^2 \right)\right) + \CO(\lam^3\rho^3)
\end{align}
while from (\ref{Hsig3}) we find that the leading correction to $f_0$ is
\begin{align}
f_1 &= 2 |mx|^2 \left( 8 \lambda^2 |m|^2 |m'|^2 c^6  -2c^2 -4 \lambda^2 |am|^2 - 2 c^2|m'x|^2 \right) +\\
&+ 2|mx|^4  \Big( \frac{2|am|^2}{c^2} + \frac{\lambda^2 |a|^4}{c^4} + 48 \lambda ^2 |m|^4 |m'|^4 c^{8}
 - 16 \lambda^2 |am|^2 |m|^2 |m'|^2 c^2   \Big) \nonumber
\end{align}
where we have again Taylor-expanded around $x=0$. 

To summarise we find that, if we add a primitive constant flux $H_1$ that commutes with the Higgs background and of the form (\ref{H1}), the D-terms equations can be solved by an appropriate choice of gauge transformation (\ref{gf}), that induces a non-primitive flux along the $\mathfrak{su} (2)$ generator $\sig_3$. When we also include the constant primitive flux $H_2$ of the form (\ref{H2})  the same is essentially true, but now we must also add a non-primitive flux $\p\bar\p h$ along the identity generator of $\mathfrak{u} (2)$ to solve the D-term constraints.

\subsection{A simple SU(3) background}
\label{s:Tbrane/SU3}

Let us now consider a slightly more complicated SU(3) T-brane background, again preserving four supercharges. The Higgs field profile in the holomorphic gauge is given by
\begin{align}
\phi_\mr{hol} = m \left(\begin{matrix}
\mu y & 1 & 0\\ ax & \mu y & 0\\
0 & 0 & -2\mu y
\end{matrix}\right)
\equiv -i m\,  E^+ + imax\, E^- + m\mu y \, Q, \label{s:gen/background}
\end{align}
where the form of the generators $E^\pm$, $Q$ and $P \equiv \left[ E^+, E^- \right]$ is detailed in Appendix \ref{ap:genAnsatz}. 

As before, we may solve for the D-terms equations by performing a gauge transformation of the form (\ref{gaugetr}). Because $[\phi, \bar\phi] \propto P$, the natural choice is now $g = \text{exp} (\frac{f}{2}P)$ and so in the unitary gauge we have a background given by
\begin{align}
\phi &= -im e^f \, E^+ + imax e^{-f} \, E^- + m\mu y \, Q \label{SU3hol}\\
F &= -i \p \pb f \, P,\nonumber
\end{align}
With this Ansatz there is only one non-trivial D-term constraint, corresponding to the generator $P$. The $\a'$ corrections complicate the form of this equation with respect to the leading order counterpart, and we obtain
\be
\left( \p_x \pb_\xb + \p_y \pb_\yb \right) f =  |m|^2 \left( e^{2f} - |ax|^2 e^{-2f} \right) \left( 1 + 4 \lambda^2 Q_f \right) -\frac{2}{3} \lambda^2R[f,f] - 4\lam^2 |m|^2 |\mu|^2 \p_x \pb_\xb f
\label{su3P}
\ee
By using the Ansatz $f=f(x,\xb)$ this expression simplifies to
\be
\p_x \pb_\xb f = \frac{|m|^2}{1+4\lambda^2 |m|^2 |\mu|^2} \left( e^{2f} - |ax|^2 e^{-2f} \right) 
\ee
which is asymptotically solved by (\ref{solf0}) with the replacement
\be
m \, \raw\, \tilde{m} = \frac{m}{\sqrt{1 + 4 \lam^2 |m|^2 |\mu|^2}}
\label{mtilde}
\ee

Let us now add further worldvolume flux to this background. For simplicity we will add it along generators that commute with the $\mathfrak{su} (2)$ subalgebra generated by $\{E^\pm, P\}$. Namely we consider the following generators
\begin{align}
T & = \left(\begin{matrix}
\mf{1}_{2 \times 2} & \\
& 0
\end{matrix}\right) \\
B &= \left(\begin{matrix}
\mf{0}_{2 \times 2} & \\
& 1
\end{matrix}\right),
\end{align}
Notice that an arbitrary combination of these generators does not belong to $\mathfrak{su} (3)$ but rather to its central extension $\mathfrak{u} (3)$. Indeed, only if we consider a worldvolume flux satisfying $F^B + 2 F^T = 0$ we will have an SU(3) background.

Similarly to the SU(2) example one may first consider a flux that commutes with the generators of the T-brane background, namely of the form
\begin{align}
\label{H1SU3}
	H_1 &= \im \left(\kappa \, dx \w d\bar{y}\right)\, T \\
	G &= M \left( \mr{d} x \w \mr{d} \xb + \mr{d}y \w \mr{d}\yb \right) \, B +  N \left( \mr{d}x \w \mr{d}\xb - \mr{d}y \w \mr{d}\yb \right)\, B + \im \left(O \, \mr{d}x \w \mr{d}\yb\right)\, B
\end{align}
where $M, N \in \IR$ and $\kappa, O \in \IC$. We may also generalise the Ansatz to $f \equiv f(x,\xb,y,\yb)$. The corrected D-term equations then read:
\begin{align}
0 &= 8 \lambda ^2 |m \mu|^2 (M+N)+N \nonumber\\
\nonumber
\left( \p_x \pb_\xb + \p_y \pb_\yb \right) f &=  |m|^2 \left( e^{2f} - |ax|^2 e^{-2f} \right) \left( 1 + \lambda^2 |\kappa|^2 + 4 \lambda^2 Q_f \right) \\ \nonumber
&-\frac{2}{3} \lambda^2R[f,f] - 4\lam^2 |m|^2 |\mu|^2 \p_x \pb_\xb f\\
0 &= \lambda^2 \re \Big( |a|^2 e^{-2 f} \kappa x \p_y f \left(2 \xb \pb_\xb f - 1 \right) + 2  e^{2 f} \kappa \p_y f \pb_\xb f \Big) \nonumber\\
& + \left( e^{2f} - |ax|^2 e^{-2f} \right) \re \Big( \kappa \p_y \pb_\xb f \Big) \nonumber\\
0 &= \lambda^2 \kappa |m \mu|^2 \left( |a|^2 e^{-2f} \left|1-2 x \p_x f \right|^2 - 4 e^{2f} |\p_x f|^2 \right)
\label{su3H1eqs}
\end{align}
Here the first equation correspond to the generator $B$ and it is identical to the D-term constraint found in (\ref{s:inter/Dterm}) for the case of intersecting 7-branes. It fixes the relation between $M$ and $N$ and decouples from the rest of the equations, that will not depend on $M, N, O$. The second equation corresponds to the D-term along the generator $P$ and it is again given by (\ref{su3P}). The third and fourth equations are new, and correspond to the D-term constraints along the generators $T$ and $E^\pm$, respectively. From the last one we see that the only way to have a non-vanishing flux $\kappa$ is to take the limit $\mu \raw 0$, which would essentially take us to the previous $SU(2)$ example.

Despite this result, one is able to accommodate a background flux along the generator $T$ by considering a slightly different Ansatz. Indeed, let us proceed as in the previous $SU(2)$ example and generalise the above flux Ansatz to
\begin{align}
	H &= H_1 + H_2 - i \p\bar{\p} h \, T \\
	H_2 &= \rho \,  i \left(dx \wedge d\bar{x} - dy \wedge d\bar{y} \right) \, T \nonumber\\
	h &\equiv h(x,\xb,y,\yb) \nonumber.
\end{align}
while returning to the Ansatz $f \equiv f(x,\xb)$ for the flux along $P$.
The corrected D-term equations now read:
\begin{align}
0 &= 8 \lambda^2 |m \mu|^2 (M+N)+N \nonumber\\
\left( 1 + 4 \lambda^2 |m \mu|^2 \right) \p_x \pb_\xb f &= |m|^2 \left( e^{2f} - |a x|^2 e^{-2f} \right)  (1 + 4 \lambda^2 Q_H ) \nonumber\\
\left( \p_x \pb_\xb + \p_y \pb_\yb \right) h &= 4 \lambda^2 |m|^2 |\mu|^2 \left(\rho -\p_x \pb_\xb h \right) \nonumber\\
&+ 2 \lambda^2 \left( \rho + \p_y \pb_\yb h \right) \left( 4 |m|^2e^{2 f} |\p_x f|^2 + 2[\phi,\phib] - |am|^2 e^{-2 f} \left|2 x \p_x f- 1 \right|^2 \right) \nonumber\\
0 &= \lambda^2 |m \mu|^2 \left( 2 \p_x \pb_\yb h + \kappa \right) \left( |a|^2 e^{-2 f} \left|2 x \p_x f - 1 \right|^2 -4  e^{2 f} |\p_x f|^2 \right) \label{Epeq}
\end{align}
with $Q_H$ again given by (\ref{QH}). Notice the last equation now imposes $2 \p_x \pb_\yb h + \kappa = 0$, which  essentially requires that the effective flux of the form (\ref{H1SU3}) vanishes. Naively, this seems to imply that $\a'$-corrected D-terms do impose constraints on worldvolume fluxes commuting with the Higgs field T-brane background, contrary to what happens at leading order in $\a'$. Nevertheless, one can show that a non-trivial $\kappa$ is allowed if one generalises the gauge transformation Ansatz $g = \text{exp} (\frac{f}{2}P)$ to include complexified transformations along the non-Cartan generators $E^\pm$ as well. We leave the somewhat technical proof of this statement  to Appendix \ref{ap:genAnsatz}, where such generalised transformations are studied in more detail. 

If for simplicity we set $\kappa=0$, make the Ansatz (\ref{expf}) and solve again perturbatively in $\lam \rho$ we find the following asymptotic solutions around $x=0$:
\begin{align}
f_0 &= \log c + c^2 |\tilde{m}x|^2 + \frac{|\tilde{m}|^2 |x|^4}{4 c^2} \left(2 c^6 |m|^2 - |a|^2 \left( 1+ 4 \lambda^2 |m \mu|^2 \right)\right) \\
f_1 &= 4 |\tilde{m} x|^2\left(2 \lambda^2 |m|^2 \left( |a|^2 - 2 c^4 \right) + c^2 \right) \\
&+ \frac{|\tilde{m}|^4 |x|^4}{c^4} \Big( \frac{|a|^2}{|m|^2} c^2 + 2 \left(\lambda ^2 \left( |a|^4 - 4 |a|^2 c^2 \left( c^2 - |\mu|^2 \right)\right)-2 c^8\right) \nonumber\\
&+8 \lambda ^2 |m|^4 \left(4 \lambda^4 |\mu|^4 \left( |a|^4 - 4 |a|^2 c^4 \right) + 2 c^6 \lambda ^2 |\mu|^2 \left( |a|^2 + 6 c^4\right)+c^{12}\right) \nonumber\\
&+4 \lambda ^2 |m|^2 \left(-4 |\mu|^2 \left(c^8-\lambda ^2 |a|^4 \right) + |a|^2 c^2 \left(4 \lambda ^2 \left( |\mu|^4 -4 c^2 |\mu|^2 \right)+c^4\right)+6 c^{10}\right) \Big) \nonumber
\end{align}
and
\begin{align}
h &= \frac{2 \lambda^2 |\tilde{m}x|^2 \rho \left(2 \left( c^4+c^2 |\mu|^2 \right)-|a|^2 \right)}{c^2} \\
&+\frac{\lambda^2 |\tilde{m} x|^4 \rho}{c^2 \left( 1 + 4 \lambda^2 |m \mu|^2 \right)} \Big(\frac{|a|^2}{|\tilde{m}|^2} \left( |m|^2 \left(3 c^2-4 \lambda^2 |\mu|^2 \right) -1 \right) +2 c^6 |m|^2 \left( \left(c^2+4 \lambda ^2 |\mu|^2 \right) +1 \right)\Big) \nonumber
\end{align}

To summarise, in this more complicated $SU(3)$ background that preserves four supercharges  we also find different kind of solutions for the $\a'$-corrected D-term equations. One first class of corrections comes from the intersection slope $\mu$ that appears in $\phi_{\rm hol}$, and which corresponds to a generator $Q$ commuting with the T-brane $\mathfrak{su} (2)$ subalgebra $\{E^\pm,P\}$. Such corrections are relatively easy to take into account, as they only modify the parameters of the Painlev\'e III equation. Further non-trivial corrections come from adding worldvolume fluxes commuting with the Higgs background. One the one hand, adding some of these primitive fluxes require a modification of the non-primitive flux $\p\pb f$ along $P$ and adding one of the form $\p\pb h$ along $T$. On the other hand, adding some other components requires a more drastic change: to generalise the standard gauge transformation $g$ to also include non-Cartan generators $E^\pm$. 
In the next section we will analyse from a more general viewpoint when each of these two cases occurs. 

\section{More general backgrounds}
\label{s:gen}

With the two examples of the previous section in mind, let us describe how $\a'$ corrections affect the D-term equations for more general kinds of T-branes. As before we will take the simplifying assumption that, given the gauge group $G$ and its corresponding Lie algebra $\mathfrak{g}$, the leading order D-term equations can be solved via a complexified gauge transformation (\ref{gaugetr}) of the form
\be
g \, =\, e^{\frac{f_i}{2} P_i}
\label{gfgen}
\ee
where $f_i= f_i(x, \xb, y, \yb)$ and $P_i$ belong to the Cartan subalgebra of $\mathfrak{g}$. We then write the Higgs field profile in the holomorphic gauge in the block diagonal form
\be
\phi_{\rm hol}\, =\, m 
\left(
\begin{array}{cccc}
\psi_{\rm hol}^1 \\
& \psi_{\rm hol}^2 \\
& &  \ddots \\
& & & \psi_{\rm hol}^n
\end{array}
\right) \, ,
\label{blocks}
\ee
with $[m] = L^{-1}$, and where the entry $\psi_{\rm hol}^i$ is an $n \times n$  matrix of holomorphic functions on $x,y$. One simple example of such structure is the SU(3) example of section \ref{s:Tbrane/SU3}, which contained a $1 \times 1$ and a $2 \times 2$ block. As discussed below eqs.(\ref{su3H1eqs}), the $\a'$-corrected D-term equations do not couple one block to the other. The same statement holds for the more general T-brane structure with the block-diagonal form (\ref{blocks}): for the purposes of analysing $\a'$-corrections we can focus on each individual block $\psi_{\rm hol}^i$ at a time, an forget about the rest. 

In the case that $\psi_{\rm hol}^i$ is a $1 \times 1$ block, the effects of $\a'$-corrections will be similar to the ones studied in section \ref{s:inter}. As in there, the $\a'$-corrections will impose primitivity with respect to the standard pull-back of $J$ on the spectral surface
\be
z \, =\, \lam\, m\, \psi_{\rm hol}^{1\times 1} (x, y) \, .
\ee

More interesting is the case where $\psi_{\rm hol}^i $ is a $2 \times 2$ block, as these contain the T-brane nature of the background. As we have already seen in section \ref{s:Tbrane} for these cases the $\a'$-corrected D-term equations may become rather involved to solve, specially when we add additional primitive worldvolume fluxes. In general, within that block we will have a holomorphic Higgs field profile of the form
\be
\psi_{\rm hol}^{2\times 2} \, =\, u_0 \mf{1} + u_1 \sig_1 + u_2 \sig_2 + u_3 \sig_3 \, =\, u_0 \mf{1} - i u_+ E^+ +  i u_- E^- + u_3 \sig_3
\label{psijhol}
\ee
where $u_i$, $u_\pm$ are complex functions on $x,y$, $[u_i] = [ u_\pm] = L^0$. Near the origin, we can approximate such functions up to their linear behaviour, so each of them is characterised by three independent complex numbers. However, we may absorb three numbers in constant shifts of the local coordinates $x,y,z$. More precisely, by a shift in $z$ we may remove the constant term in $u_0$, rendering it a linear function in $x,y$. Similarly, by shifts on $x$ and $y$ we may remove the constant pieces in $u_3$ and $u_-$. Then we are left with only one function, namely $u_-$ that may contain a constant term, and therefore with essentially two different possibilities
\be
\left.\psi_{\rm hol}^{2 \times 2}\right|_{x=y=0}  = 
\left(
\begin{array}{cc}
 0& 1 \\ 0 &0
\end{array}
\right) \qquad \text{and} \qquad
\left.\psi_{\rm hol}^{2 \times 2}\right|_{x=y=0}  =  
\left(
\begin{array}{cc}
 0& 0 \\ 0 &0
\end{array}
\right) \, .
\label{epsilon}
\ee
Examples of backgrounds of the first kind are those analysed in section \ref{s:Tbrane}, while several of the second kind are studied in Appendix \ref{ap:more}. In both cases the holomorphic Higgs background is parameterised by eight dimension-full parameters, namely
\be
\begin{array}{lcl}
u_0 \, =\, \mu_{0, x} x + \mu_{0, y} y & & u_3 \, =\, \mu_{3, x} x + \mu_{3, y} y\\
u_- \, =\, \mu_{-, x} x + \mu_{-, y} y & & u_+ \, =\, \mu_{+, x} x + \mu_{+, y} y + \eps
\end{array}
\ee
where $[\mu_{i,\a}] = L^{-1}$ and $\eps = 0, 1$ describes the two cases in  (\ref{epsilon}). Imposing that the leading order D-term equation  is solved by  (\ref{gfgen}) means that at $\lam \raw 0$ we need a complexified gauge transformation of the form
\be
g^{2\times 2}\, =\, e^{\oh (f \sig_3 + h \mf{1}_2)}
\label{gf2x2}
\ee
for solving the $2 \times 2$ block which we are analysing. In practice, this is only possible if $[\psi_{\rm hol}^{2\times 2},\conj{\psi_{\rm hol}^{2\times 2}}] \in \text{Cartan}$, which requires $\mu_{3,x} = \mu_{3,y} = 0$. We then have that in our setup
\begin{align}
	\psi_{\rm hol}^{2\times 2} &= \left( \begin{matrix}
	\mu_{0,x} x + \mu_{0,y} y & \mu_{+,x} x + \mu_{+,y} y + \epsilon \\
	\mu_{-,x} x + \mu_{-,y} y & \mu_{0,x} x + \mu_{0,y} y
	\end{matrix} \right).
\end{align}

One may now wonder if taking into account $\a'$-corrections will drastically change the form of the complexified gauge transformation (\ref{gf2x2}) solving for the D-term equation. For this we observe that
\begin{itemize}

\item If no background fluxes along $\mf{1}_2$ are present, then the Ansatz (\ref{gf2x2}) remains invariant (with $h\equiv 0$), although $\a'$-corrections may vary the specific form of $f$ with respect to its leading order value.

\item If we switch a background flux $H$ along $\mf{1}_2$ then, for a generic $\psi^{2\times 2}_{\rm hol}$, some components of $H$ will preserve the Ansatz (\ref{gf2x2}), while others will force to consider a gauge transformation including non-Cartan generators $E^\pm$, as discussed in Appendix \ref{ap:genAnsatz}.

\end{itemize}

Let us be more precise on the last point, since adding non-Cartan generators to (\ref{gf2x2}) implies having a non-Abelian flux background that will complicate the T-brane system. By inspection (see e.g., Appendix \ref{ap:genAnsatz}) one quickly realises that the relevant D-term equations for this problem are those along the non-Cartan components $E^\pm$, which may or may not have solution for the Ansatz (\ref{gf2x2}). If there is no solution, one needs to generalise this Ansatz to include the generators $E^\pm$ and therefore a non-Abelian gauge background appears through (\ref{gaugetr}).

Due to the symmetrisation procedure, the D-term equations along $E^\pm$ receive contributions only from the middle term in eq.\eqref{s:Dterms/D-term}. More precisely, assuming the Ansatz (\ref{gf2x2}) we have that
\begin{align}
	\D \psi^{2\times 2} &\equiv (\D \psi)_{\mf{1}} \, \mf{1}_2 + (\D \psi)_+ E^+ + (\D \psi)_- E^-  \\
	&=\left( \mu_{0,x} \mr{d} x + \mu_{0,y} \mr{d} y \right) \, \mf{1}_2 \nonumber \\
	&+\left( \mu_{+,x} \mr{d} x + \mu_{+,y} \mr{d} y + 2 \p f \left( \mu_{+,x} x + \mu_{+,y} y + \epsilon \right) \right) \, e^f E^+ \nonumber\\
	&+\left( \mu_{-,x} \mr{d} x + \mu_{-,y} \mr{d} y - 2 \p f \left( \mu_{-,x} x + \mu_{-,y} y \right) \right) \, e^{-f} E^- \nonumber ,
\end{align}
and that the D-term equations along $E^\pm$ read
\begin{align}
0 &= D_\pm = 2i \lambda^2 \Big( (\D \psi)_\pm \w \overline{(\D \psi)_{\mf{1}}} + (\D \psi)_{\mf{1}} \w \overline{(\D \psi)_\mp} \Big) \w H\, .
\label{DEpm}
\end{align}
From here we see that these equations are non-trivial only if the Higgs-vev $\psi^{2 \times 2}_{\rm hol}$ has components simultaneously along the identity and a (non-Cartan) generator of $\mathfrak{su} (2)$, which will be generically the case. Moreover, the total background flux $H$ along the identity (including the piece $-i \p\bar{\p} h$) must be non-vanishing for this equation to be non-trivial. Let us discuss how this condition constrains the background flux $H$. Recall that $H$ must satisfy the corrected primitivity condition
\begin{align}
\label{primiH}
	0& = \omega \wedge H 
	+ \lam^2 \left( 2i (\D \psi)_{\mf{1}} \w \overline{(\D \psi)_{\mf{1}}}  - 2 \im [(\D \psi)_+ \w \overline{(\D \psi)_-}]  - \tr ([\phi,\phib] F) \right) \w H 
\end{align}
and satisfy the Bianchi identity $dH= 0$. Then we find that only some profiles for $H$ may satisfy the complex equations (\ref{DEpm}) and the real equation (\ref{primiH}) simultaneously. Those profiles that satisfy (\ref{primiH}) but fail to satisfy (\ref{DEpm}) will not be compatible with the initial Ansatz (\ref{gf2x2}) and therefore will require the presence of a non-Cartan flux background at $\CO(\lam^2)$. 

In practice one may find by inspection which profiles for $H$ are compatible with the Abelian Ansatz (\ref{gf2x2}), although in some simple cases one may be more specific. In particular, let us consider the cases where

\begin{itemize}

\item{$(\D \psi)_+  \wedge (\D \psi)_- = 0$}

Or equivalently  $(\D \psi)_+ = \gamma (\D \psi)_-$ for some complex function $\gamma$. In this case one finds that all fluxes $H$ of the form
\bea
\label{pm1}
H  & \propto & i (\D \psi)_{\mf{1}} \w \overline{(\D {\psi})_{\mf{1}}}\\
H  & \propto & i (\D \psi)_- \w \conj{(\D \psi)_-}
\label{pm2}
\eea
satisfy eq.\eqref{DEpm}.
Moreover if $\bar \g \equiv \g^{-1} $ then both equations in \eqref{DEpm} become the same. In particular for $\g \equiv \eta = \pm 1$ they become a real condition and 
\be
H \, \propto \, \re \left[ \sqrt{\eta}\, (\D \psi)_- \w \conj{(\D \psi)_\mf{1}} \right] 
\label{pm3}
\ee
also becomes a solution to \eqref{DEpm}. Any combination of these allowed components satisfying $dH=0$ and (\ref{primiH}) will not require a non-Abelian flux background, while the rest will. 

\item{$(\D \psi)_\pm  \wedge (\D \psi)_\mf{1} = 0$}

Or equivalently $(\D \psi)_\pm = \gamma (\D \psi)_\mf{1}$ for a complex function $\g$. In this case again both equations in \eqref{DEpm} becomes conjugate to each other and
\bea
H  & \propto & i (\D \psi)_{\mf{1}} \w \overline{(\D {\psi})_{\mf{1}}}\\
H  & \propto & \im \left[ \g (\D \psi)_\mp \w \conj{(\D \psi)_\mf{1}} \right] + \frac{i}{2} (\D \psi)_\mp \w \conj{(\D \psi)_\mp}
\eea
automatically satisfy \eqref{DEpm}. Again, a combination of those satisfying (\ref{primiH}) and $dH=0$ will be compatible with an Abelian flux background.

\item{$(\D \psi)_+  \wedge (\D \psi)_- =  (\D \psi)_+  \wedge (\D \psi)_\mf{1} = (\D \psi)_-  \wedge (\D \psi)_\mf{1} =0$}

In this case we have that  \eqref{DEpm} will be solved by
\bea
H  & \propto & i (\D \psi)_{\mf{1}} \w \overline{(\D {\psi})_{\mf{1}}}\\
H  & \propto & \im \left[ \g (\D \psi)_{\mf{1}} \w \conj{\eta} \right]
\eea
for arbitrary complex function $\g$ and one-form $\eta \in \Omega^{(1,0)}$. Such that we have more freedom to satisfy primitivity condition and Bianchi identity than in the previous cases.

\end{itemize}

One can check that this general discussion reproduces the results found in the two simple examples of section \ref{s:Tbrane}. On the one hand, for the $SU(2)$ example of section \ref{s:Tbrane/SU2} we have that $(\D \psi)_{\mf{1}} =0$. Hence \eqref{DEpm} is trivially satisfied and so non-Cartan fluxes are absent in the corrected solution. On the other hand, in the $SU(3)$ example of section \ref{s:Tbrane/SU3}, the $2 \times 2$ T-brane block is such that
\be
(\D \psi)_+, (\D \psi)_- \propto dx\, , \qquad\qquad (\D \psi)_{\mf{1}} \propto dy\,
\ee
We are then in the case $(\D \psi)_+ = \gamma (\D \psi)_-$, with $\g$ a complicated function. It is then easy to see that 
\be
H\, =\,  \rho \,  i \left(dx \wedge d\bar{x} - dy \wedge d\bar{y} \right) + \CO(\lam^2) , \quad \qquad \rho \in \IR
\ee
is a linear combination of the two-forms (\ref{pm1}) and (\ref{pm2}) which satisfies the Bianchi identity and the primitivity condition at leading order. This is precisely the flux component denoted as $H_2$ in section \ref{s:Tbrane/SU3}, explicitly shown to be compatible with the Abelian Ansatz (\ref{gf2x2}) therein. On the contrary, a flux of the form (\ref{H1SU3}) is shown to be incompatible with such an Ansatz, and non-Cartan flux generators need to be added as described in Appendix \ref{ap:genAnsatz}. This again matches our general discussion, as for some choices of $\kappa$ the flux (\ref{H1SU3}) can be made of the form (\ref{pm3}). But since in this example $\gamma \neq \pm 1$ such a flux is incompatible with the naive Abelian Ansatz, and non-Cartan generators need to be included.

\section{Applications to local F-theory models}
\label{s:app}

The T-brane backgrounds that we considered in the previous section are very similar to those used to generate phenomenological Yukawa hierarchies in F-theory GUTs \cite{Font:2013ida,Marchesano:2015dfa,Carta:2015eoh}, with the main difference that there $\Phi$ and $F$ are valued in the Lie algebra of the exceptional groups $E_6, E_7$ and $E_8$. Nevertheless, in order to build models of SU(5) unification the Higgs background is embedded in unitary subalgebras of these exceptional groups and, at least naively, one may use this fact to apply our results. 

Let us for instance consider the $E_6$ T-brane background constructed in \cite{Font:2013ida}
\begin{align}
	\phi = m \left( e^f E^+ + mx e^{-f} E^- \right) + \mu^2 (bx-y) Q \, ,
	\label{e6Higgs}
\end{align}
where the generators $E^\pm$ generate a $\mathfrak{su} (2)$ subalgebra via $[E^+, E^-] = P$ and $Q$ a commuting $\mathfrak{u} (1)$ subalgebra, see \cite{Font:2013ida} for precise definitions. This background is quite similar to the one considered in section \ref{s:Tbrane/SU3}, as one can see from acting with $\phi$ on the doublet sector $(\bf{10}, \bf{2})_{-1}$  within the adjoint of $\mathfrak{e}_6$ \cite{Font:2013ida} 
\be\label{com10}
[\phi, R_+E_{\mathbf {10}^+}+R_-E_{\mathbf {10}^-}] = 
\left(
\begin{array}{cc}
-\mu^2(bx-y) & m \\ m^2 x & -\mu^2(bx-y)
\end{array}
\right)
\left(
\begin{array}{c}
R_+ E_{\mathbf {10}^+} \\ R_-E_{\mathbf {10}^-}
\end{array}
\right)\, .
\ee
Naively, this action can be identified with a $2 \times 2$ Higgs block $\psi^{2 \times 2}$ of the sort discussed in section \ref{s:gen}. In fact, it is identical to the $2 \times 2$ block that arises from eq.(\ref{SU3hol}) if there we perform the replacements
\be
y\, \raw\, y - bx\, , \qquad \qquad a \, \raw \, m\, , \qquad \qquad m \mu \, \raw \, \mu^2 \, .
\ee

One can now apply the analysis of the previous section to this case. As in the SU(3) example of section  \ref{s:Tbrane/SU3}, we are in the case $(\D \psi)_+ = \gamma (\D \psi)_-$ for $\gamma \neq \pm 1$. Therefore, primitive fluxes of the kind $H_\text{nc} \mf{1}_{2\times 2}$ with a component of the form
\begin{align}
	H_\text{nc} \propto \mr{Re} \big( (\D \psi)_- \w \conj{(\D \psi)_\mf{1}} \big) \propto \mr{Re} \big( \mr{d}x \w ( \conj{b} \mr{d} \xb - \mr{d} \yb )  \big)
\end{align}
are not allowed at order $\lam^2$ without adding further non-Cartan fluxes. Interestingly, for the case $b=1$ used in \cite{Font:2013ida} to compute physical Yukawas, we have that such problematic flux reads
\be
H_\text{nc} \stackrel{b=1}{\propto}  \mr{Re} \big( \mr{d}x \w \mr{d} \yb ) \, ,
\ee
which allows for some primitive fluxes. In fact, the worldvolume primitive fluxes considered in \cite{Font:2013ida} were of the form 
\be
F_p \, =\, i Q_R (dy \w d\yb - dx \w d \xb) + i Q_S (dx \w d\yb + dy \w d\xb)
\label{Fp}
\ee
with $Q_R$, $Q_S$ some Cartan generators that reduce to the identity for the sector of interest. Therefore, according to our naive analysis the presence of these primitive fluxes may modify the non-primitive Abelian flux Ansatz given by $g = \text{exp}(\oh fP)$ with $f= f(x, \xb)$, but it will not require the presence of non-Cartan generators in the flux background. Hence it seems that the computation of physical Yukawas made in \cite{Font:2013ida} may be affected by $\alpha'$ corrections but not drastically, in the sense that the Ansatz for the T-brane background taken there survives at the next-to-leading order in $\a'$. This will change as soon as the worldvolume flux (\ref{Fp}) is chosen more general or $b$ is chosen such that $\im\, b \neq 0$.

\section{Conclusions}
\label{s:conclu}

In this paper we have analysed the effect of $\a'$-corrections on BPS systems of multiple D7-branes, with special emphasis on T-brane configurations. Our main strategy has been to compute how $\a'$-correction modify the D-term BPS condition, solve for the new background profiles for $\Phi$ and $A$, and compare them with the previous leading-order D-term solution. Since $\a'$-corrections do not enter holomorphic D7-brane data, this comparison can be made in terms of the complexified gauge transformation (\ref{gaugetr}) in terms of which we solve the D-term equations. 

In D7-brane T-brane systems, solving the D-term equation is quite involved already at leading order, which renders our analysis somewhat technical. Nevertheless, we have drawn several lessons from the cases that we have analysed: 

\begin{itemize}

\item When the Higgs background takes a block-diagonal form (\ref{blocks}), $\a'$-corrections can be analysed block by block, as they do not couple different blocks. 

\item For system of intersecting D7-branes $\a'$-corrections have a simple interpretation in terms of the pull-back of the K\"ahler form on the actual D7-brane embedding. It would be interesting to see if T-brane systems allow for a similar interpretation.  

\item In all the examples that preserve eight supercharges, $\a'$-corrections do not modify the background. The classical solution also solves the corrected D-term equations. A trivial example of this are intersecting D7-branes without fluxes.

\item One may lower the amount of supersymmetry to four supercharges by modifying the Higgs field by a constant slope $\Delta\Phi$ or by adding a constant primitive flux $H$, both commuting with the group generators involved the T-brane background. At leading order these additions do not modify the T-brane background at all. When $\a'$-corrections are taken into account the T-brane background is modified, but there are several degrees of complexity at which this may happen

\begin{itemize}

\item[{\em i)}] 
In the simplest case $\a'$-corrections only modify the dimensionful parameters which enter the differential equation for the non-primitive flux background (\ref{npflux}) and the related complexified gauge transformation (\ref{gaugetr}), as in eqs.(\ref{mprime}) and (\ref{mtilde}). Hence they can be typically absorbed into a coordinate redefinition.

\item[{\em ii)}] 
In slightly more complicated cases we need to generalise the complexified gauge transformation to
\be
g \, =\, e^{\oh (f P + h \mf{1})}
\label{Abg}
\ee
to absorb the effect of some primitive flux $H$. The corresponding non-primitive flux is therefore still Abelian, with $f$ being modified from the leading-order expression. The equations governing $f$ and $h$ are rather complicated, but one may solve them by performing a perturbative expansion in $\a'$-suppressed parameters. More precisely we have assumed the following hierarchy
\be
 \a'  \rho_i \ll \a' m_j^2 \ll 1
 \label{hlimit}
\ee
to find solutions to next-to-leading order in $\a'$. Here $\rho_i$ are primitive flux density parameters and $m_j$ T-brane slope parameters.  

\item[{\em iii)}] 
In the most complex case the Abelian Ansatz (\ref{Abg}) is not sufficient to solve the corrected D-term equations, which develop non-trivial components along non-Cartan generators (in particular those which the holomorphic T-brane data depends on). One then needs to consider a complexified gauge transformation that depends on such generators, as in Appendix \ref{ap:genAnsatz}. The analysis for these corrected backgrounds is even more involved and one again needs to resort to a perturbative expansion to find solutions to next-to-leading order in $\a'$.

\end{itemize}

\item This last, more complicated case contains all the ingredients that are generic in the construction of 4d chiral local F-theory GUT models, so one may speculate that $\a'$-corrections could change qualitatively the description of these configurations, as we have briefly discussed. In any event, the holomorphic data of these models will not be affected by $\a'$-corrections. In particular the holomorphic Yukawa hierarchies of \cite{Font:2013ida,Marchesano:2015dfa,Carta:2015eoh}, which only depend on such holomorphic data, will still be present after $\a'$-corrections are taken into account.

\end{itemize}

Based on these results, one may conceive of several directions to pursue the analysis of $\a'$-corrections in T-brane systems. First, it would be interesting to extend our background solutions to higher orders in the $\a'$ expansion and beyond the limit (\ref{hlimit}). Second, it would be interesting to see if the interpretation of $\a'$-corrections for the intersecting D-brane case can be incorporated in some form for T-brane backgrounds. Moreover, it would be interesting to verify our naive analysis of $\a'$-corrections in F-theory local models based in exceptional groups, and compute how $\a'$-corrections modify the normalisation of chiral mode wavefunctions in realistic models. Finally, it would be interesting to see the consequences of our findings for the recent proposal to use T-branes in the construction of de Sitter vacua  \cite{Cicoli:2015ylx}.

\bigskip

\bigskip

\centerline{\bf \large Acknowledgments}

\bigskip

We thank M.~Montero, D.~Regalado, R.~Savelli, A.~Uranga and G.~Zoccarato for useful discussions. 
This work has been partially supported by the grants FPA2012-32828 and FPA2015-65480-P from  MINECO, SEV-2012-0249 of the ``Centro de Excelencia Severo Ochoa" Programme, and the ERC Advanced Grant SPLE under contract ERC-2012-ADG-20120216-320421. S.S. is supported through the FPI grant SVP-2014-068525.


\appendix


\section{D-terms from the Chern-Simons action}
\label{ap:CS}

In section \ref{s:Dterms} we discussed how to derive the D-terms for non-Abelian stacks of D7-branes in IIB orientifolds with O3/7-planes via their generalised calibration conditions. As we will now show, one can reach the same result by considering the 4d couplings that arise from the Chern-Simons action. Indeed, as was argued in \cite{Dine:1987xk}, the D-terms of the four dimensional effective action are related by supersymmetry to terms of the form $\int \tilde{B}_2 \w F$, where $\tilde{B}_2$ is a 4d two-form dual to an axion and $F$ the field strength of a gauge group generator. As in other D-brane setups here the two-forms $\tilde{B}_2$ arise from RR $p$-forms, and so such couplings will be contained in the D-brane Chern-Simons action.

The non-Abelian Chern-Simons action for a stack of D7-branes is given by \cite{Myers:1999ps}
\begin{align}
S_\mr{CS} &= \mu_p \int_{\IR^{1,3} \times \mc{S}}  \hspace*{-.4cm}\mr{STr} \left( \mr{P} \left[ e^{i \lambda \iota_\Phi \iota_\Phi} \sum C^{(n)} \w e^{-B} \right] \w e^{\lambda F} \right), \label{ap:CS/CS}
\end{align}
where we will use the same parametrisation for the Higgs-field as in the main text
\begin{align}
\Phi &= \phi \frac{\p}{\p z} + \phib \frac{\p}{\p \zb}.
\end{align}
For simplicity, let us assume that the odd cohomology groups of the compactification manifold $H^2_-(X_3) = H^4_-(X_3)$ vanish. Then  the harmonic components of the internal B-field are projected out, and the same applies to the 4d two-forms that could arise from the dimensional reduction of the RR forms $C_2$ and $C_6$. The only relevant 4d two-forms and their axion duals arise from the expansion of the orientifold-even RR forms
\begin{align}
C^{(4)} &  = \, c_2^a \omega_a + \rho_a \tilde\omega^a + \dots \\
C^{(8)} & =\, e_2 \, \omega_6 + \dots
\end{align}
where $\omega_a$, $\tilde\omega^a$ run over the bases of integer two- and four-forms in the internal space, respectively (such that $J = e^{\phi_{10}/2} v^a \omega_a$) and $ \omega_6 = d\mr{vol}_X/\sqrt{g_X}$ is the unique harmonic six-form with unit integral over $X_3$. Plugging this into \eqref{ap:CS/CS} gives
\begin{align}
S_\mr{CS} &\supset \lambda^2  \mu_p \int_{\IR^{1,3} \times \mc{S}} \hspace*{-.4cm} \mr{STr} \Bigg\{ F_{4d} \wedge \left[e_2 \w i \iota_\Phi \iota_\Phi \omega_6 
+c_{2}^a \w \bigg( P[\omega_a] \w F + \frac{i \lambda^2}{2} \iota_\Phi \iota_\Phi \left(\omega_a \right) F^2  \bigg)  \right]
\Bigg\},
\label{ap:CS/CS2}
\end{align}
where $F_{\rm 4d}$ stands for the components of the D7-brane field strength with legs on $\IR^{1,3}$, and we have imposed the absence of internal B-field. 

The two-forms coupling to $F_{\rm 4d}$ have as 4d duals
\be
dc_2^a = \frac{g^{ab}}{4\CK^2} *_{\IR^{1,3}} d \rho_b \qquad \qquad de_2 =  e^{2\phi_{10}}*_{\IR^{1,3}} d C_0
\ee
where $\tau = C_0 + i e^{-\phi_{10}}$ is the type IIB axio-dilaton, $\CK = \frac{1}{6} \CK_{abc} v^a v^bv^c$ with $\CK_{abc}$ the triple intersection numbers of $X_3$, and $g^{ab}$ is the inverse of $g_{ab} = \frac{1}{4\CK}\int_{X_3} \om_a \wedge * \om_b$ . Such duality relations tells us how a vector multiplet coupling to $c_2^a$ and $e_2$ enters the type IIB K\"ahler potential. Let us start from the usual expression
\be
K_{\rm IIB} \, =\, - \text{log} (S + \bar{S})  - \text{log} (\CK^2) - \text{log} \left( \int \Om \wedge \bar{\Om} \right)
\label{KIIB}
\ee
where $S = -i \tau$. Here $\CK^2$ should be seen as a function of $\re\, T_a$, with $T_a = - \oh \CK_{abc} v^a v^b - i \rho_a$. Then a vector multiplet $V_i$ coupling to these axions via a St\"uckelberg coupling $Q^i_\a$ should enter the K\"ahler potential (\ref{KIIB}) through the replacements
\be
S + \bar{S} \  \raw\ S + \bar{S} - Q_0^i V_i\, , \qquad \qquad T_a + \bar{T}_a\ \raw \ T_a + \bar{T}_a - Q_a^i V_i\, .
\ee
Finally, the Fayet-Iliopoulos term corresponding to $V_i$ will be given by
\be
\xi_i\, \propto\, \left(\frac{\p K}{\p V_i}\right)_{V=0}\, .
\ee
This prescription has been applied in \cite{Cremades:2002te} to reproduce the D-terms of intersecting D6-brane models, which automatically include the $\a'$ corrections of mirror type IIB setups. The latter have been analysed from this viewpoint in the Abelian case in  \cite{Plauschinn:2008yd}. In the following we will see that it can also be used to reproduce the D-terms of $\a'$-corrected non-Abelian D7-brane systems.

Indeed, we may apply the above prescription generator by generator of the non-Abelian gauge group of the D7-brane stack, extracting the St\"uckelberg charges $Q^i_\a$ from the couplings $\int_{\IR^{1,3}} \tilde{C}_2^\a \wedge F_i$. At the end we obtain that the above prescription amounts to perform the following replacement in (\ref{ap:CS/CS2})
\be
e_2 \ \raw \ e^{\phi_{10}}\, , \qquad \qquad c_2^a \ \raw \ - \frac{v^a}{\CK} \, ,
\ee
 that is, to trade the two forms by their partners in the corresponding linear multiplet. We then finally obtain a non-Abelian D-term proportional to
%
%
%
\begin{align}
\lambda^2 \mu_p \int_\mc{S} \mr{S} \Bigg\{P[J] \w F + \frac{i \lambda^2}{2} (\iota_\Phi \iota_\Phi J) {F}^2 
 -\frac{i}{6} \iota_\Phi \iota_\Phi J^3 \nonumber
\Bigg\}.
\end{align}
where we have used that $J = e^{\phi_{10}/2} v^a \omega_a$. Hence we precisely recover the expression as in \eqref{s:Dterms/D-term}. Finally, a similar analysis can be done for the case of non-vanishing internal B-field to recover \eqref{s:Dterms/D-termB}.

\section{Globally nilpotent T-brane backgrounds}
\label{ap:toda}
In \cite{Bena:2016oqr} it was recently shown that certain non-Abelian D7-brane vacuum solutions may be described in terms of a single curved D7-brane. More specifically, these vacua are compactifications of IIB string theory on $\mb{R}^{1,5} \times \mb{C}^2$ with a globally nilpotent Higgs-vev in $SU(N)$. Taking $(x,z)$ to parametrise the $\mb{C}^2$-factor, the D7-brane stack on $\{z=0\}$ is described by
\be
\Phi=\left(\begin{array}{cccccc} 0&\phi_1& & &\\ &0 & \phi_2 & & 0 \\ & & \ddots&\ddots& \\ &0&&0&\phi_{N-1}\\&&&&0 \end{array}\right)\,,\qquad \phi_a=\sqrt{a(N-a)}\,e^{C_{ab}f_b/2}\,,
\ee
where $C_{ab}$ is the Cartan matrix of $SU(N)$ and the $\{f_a\}$ are functions of the D7-brane world-volume coordinates $(x,\xb)$. The flux is given as
\begin{align}
	F = - \p \pb f_a C_a,
\end{align}
where the $C_a$ are the Cartan generators of $SU(N)$. In this reference, explicit solutions $\{f_a\}$ to the D-term equations have been computed at leading order in $\alpha'$. This leading order solution was then used to provide a description of this system in terms of a single, curved D7-brane. The latter description is in principle valid whenever the field vevs are large compared to $\alpha'$, but the authors of \cite{Bena:2016oqr} noted that their solution should also be valid in regions where such vevs are small, due to the characteristic of their solution. 

In the following we will take a complementary viewpoint and analyse the above background via the non-Abelian Hitchin system, better suited for for small field vevs. We will compute their $\alpha'$-corrections explicitly and see that, just like in other T-brane backgrounds preserving eight supercharges, the classical solution is still valid after $\a'$-corrections are taken into account. This implies that the classical  analysis encodes all the information of the system, and that the dictionary built in \cite{Bena:2016oqr} is not affected by $\a'$-corrections.

Indeed, from eq.(\ref{s:Dterms/D-term}), we know that the corrected D-term equations are of the form $D = D_0 + \lambda^2 D_2 = 0$, with $D_0$ the leading order D-term and $D_2$ given by
\begin{align}
D_2 = \int_{\mc{S}} \mr{S} &\Bigg\{ 2i \D \phi \w \conj{\D} \phib \w F - \left[ \phi, \phib \right] F^2 \Bigg\}.
\end{align}
However in this background $F$, $\D \phi$ and $\Db \phib$ only have legs along $\mr{d}x$ and $\mr{d} \xb$, and therefore $D_2$ vanishes identically. Hence, the whole system is insensitive to $\a'$-corrections irrespective of how large the values for $\langle \phi \rangle$, $\langle D \phi \rangle$ and $\langle F \rangle$ are. 

\section{Non-cartan flux backgrounds}
\label{ap:genAnsatz}

When analysing non-Abelian D-term equations in section \ref{s:Tbrane}, we have always made the Ansatz that the gauge transformation $g$ that defines the non-primitive flux lies entirely within the Cartan subalgebra of the gauge group $G$. However, when analysing $\a'$-corrected D-terms, the gauge derivatives generically introduce contributions to the D-terms also along the non-Cartan generators. Hence, it is natural to wonder whether adding worldvolume fluxes along non-Cartan generators may provide new solutions to the D-term equations. 

In general, introducing non-Cartan fluxes via a gauge transformation leads to very involved BPS equations. For the setup at hand we may, however, follow a simple approach. Since we know that at leading order in $\lambda$ no such flux is required to solve the D-term equations, we may assume that it is purely a $\lambda$-correction. This suggests that we capture the relevant physics if we perform an infinitesimal gauge transformation
\begin{align}
	\phi &\longrightarrow \phi + [\delta g, \phi]\\
	\conj{A} &\longrightarrow \conj{A} + i \pb \delta g,
\end{align}
with $\delta g$ proportional to some small parameter $\lam^2 \a$, $[\a] = L^{-4}$. In the following we will implement this strategy  for the two T-brane backgrounds analysed in section \ref{s:Tbrane}.

\subsubsection*{SU(2) example}

Let us consider the $SU(2)$ background analysed in subsection \ref{s:Tbrane/SU2}, which we reproduce here for convenience
\begin{align}
	\phi &= m \left(\begin{matrix} 0 & e^f\\ a x e^{-f} & 0 \end{matrix}\right)\, ,\\
	F &= -i \p \pb f \, \sigma_3 -i \p \pb h \, \mf{1} \, .
\end{align}
On top of this background we perform a gauge transformation of the form 
\begin{align}
	\delta g &\equiv \lambda^2 \left( \frac{\alpha}{2} E^+ + \frac{\conj{\alpha}}{2} E^- \right)\, ,
\end{align}
where 
\be
E^+\, =\, 
\left(
\begin{array}{cc}
0 & i \\ 0 & 0
\end{array}
\right) \, ,\qquad \qquad
E^-\, =\, 
\left(
\begin{array}{cc}
0 & 0 \\ -i & 0
\end{array}
\right) \, .
\ee
Notice that the relation between the gauge parameters multiplying $E^\pm$ is necessary for the resulting flux to satisfy the Bianchi identity. Acting on the above background such gauge transformation gives
\begin{align}
	\delta \phi &= - \frac{i \lambda^2 m}{2} \left( \alpha a x e^{-f} + \conj{\alpha} e^f \right) \sigma_3 \\
	\delta F &= -i \lambda^2 \p \pb \left( \alpha \, E^+ - \conj{\alpha} \, E^- \right).
\end{align}
We then plug this into the D-term equations and consider the linear terms induced by this infinitesimal transformation
\begin{align}\label{SU2inffluct}
	&\omega \w \delta F + \omega^2 \left( [\phi, \conj{\delta \phi}] + [\delta \phi, \phib ] \right)
	= \, \frac{\lambda^2}{2} \left( \p_x \pb_\xb + \p_y \pb_\yb \right) \left( \alpha \, E^+ + \conj{\alpha} \, E^- \right) + \\
	&+\frac{\lambda^2 |m|^2}{2} \left( 2 \conj{\alpha} \conj{a} \xb + \alpha e^{2f} + \alpha |ax|^2 e^{-2f} \right) E^+ 
	+\frac{\lambda^2 |m|^2}{2} \left( 2 \alpha a x + \conj{\alpha} e^{2f} + \conj{\alpha} |ax|^2 e^{-2f} \right) E^-. 
	\nonumber
\end{align}
Interestingly the infinitesimal gauge transformation only introduces components in $E^\pm$, which means these new contributions are entirely decoupled from the D-term equations within the main text and may be considered independently.

From \eqref{SU2inffluct} we read off, that the parts in $E^+$ and $E^-$ are conjugate to each other, and so we only need to satisfy one new D-term equation:
\begin{align}
	\left( \p_x \pb_\xb + \p_y \pb_\yb \right) \alpha &= - 2 \conj{\alpha} \conj{a} \xb |m|^2 - \alpha |m|^2 \left( e^{2f} + |ax|^2 e^{-2f} \right).
\end{align}
which we may solve asymptotically near the origin by plugging in the solution for $f$ given in \eqref{solf0}
\begin{align}
	\alpha &= \gamma \left( 1 - c^2 |mx|^2 - \frac{|mx|^4}{4c^2} \left( c^6 + \frac{|a|^2}{|m|^2} \right) \right)\, , 
\end{align}
where $\g \in \IC$ and $[\g] = L^{-4}$.
We may interpret this one-parameter solution as a massless deformation to the T-brane background allowed at the infinitesimal level by the F- and D-terms. As pointed out in \cite{Cecotti:2010bp}, this $SU(2)$ background contains one zero mode precisely along the generators $E^\pm$. Therefore it is natural to relate the parameter $\g$ with the vev of this zero mode.

\subsubsection*{SU(3) example}

Let us now apply this strategy to the $SU(3)$ background of subsection \ref{s:Tbrane/SU3}, more precisely we act with the infinitesimal gauge transformation
\begin{align}
	\delta g &\equiv \lambda^2 \left( \frac{\alpha}{2} E^+ + \frac{\conj{\alpha}}{2} E^- \right),
\end{align}
on the background \eqref{SU3hol}. Now
\be
E^+\, =\, 
\left(
\begin{array}{ccc}
0 & i & 0\\ 0 & 0 &0 \\ 0 & 0 &0 
\end{array}
\right) \, ,\qquad 
E^-\, =\, 
\left(
\begin{array}{ccc}
0 & 0 & 0 \\ -i & 0 & 0 \\ 0 & 0 &0 
\end{array}
\right) \, ,\qquad 
P\, =\, 
\left(
\begin{array}{ccc}
1 & 0 & 0 \\ 0 & -1 & 0 \\ 0 & 0 &0 
\end{array}
\right) \,
\ee
so this transformation takes us to
\begin{align}
	\tilde{\phi} &= \phi + \frac{\lambda^2}{2} m \left( \alpha m x e^{-f} - \conj{\alpha} e^f \right) P \\
	\conj{\tilde{A}} &= \Ab + \frac{i \lambda^2}{2} \pb \left( \alpha E^+ + \conj{\alpha} E^- \right)\, ,
\end{align}
so that we get new contributions to the D-term equations given by
\begin{align}
	\delta D &= -i \lambda^2 \omega \w \p \pb \alpha\, E^+ + \lambda^2 \conj{m} \xb |m|^2 \conj{\alpha}\, E^+ - \frac{\lambda^2}{2} \alpha |m|^2 \left( e^{2f} + |mx|^2 e^{-2f} \right)\, E^+ + \text{h.c.}
\end{align}
again exclusively along the non-Cartan generators $E^\pm$. This time the D-term equations have already some components along such non-Cartan generators.\footnote{More precisely, the fourth equation in \eqref{Epeq} is a linear combination of those in the generators $E^+$ and $E^-$ --- which are conjugate to each other. The equation in $E^+$ reads \begin{align}
	D^+ &= -i \lambda^2 |m|^2 \left(2 \conj{\mu} e^{f} \p_x f \left(2 \p_y \pb_\xb h + \conj{\kappa} \right) - \mu \conj{a} e^{-f} \left(2 \xb \pb_\xb f - 1 \right) \left(2 \p_x \pb_\yb h + \kappa \right)\right)
\end{align}}
Recall from the discussion in the main text that it is precisely this equation that forced to set $\kappa = 0$. Therefore one may wonder if these new contributions proportional to $\a$ may allow for a non-trivial $\kappa$. 
Indeed, one can confirm that a gauge transformation given by
\begin{align}
	\alpha &= \xb \alpha_0 + |x|^2 \alpha_1 + \xb |x|^2 \alpha_2 + \dots,
\end{align}
where the constant coefficients $\alpha_i$ depend intricately on $\kappa, f, \dots$ is such a solution. For instance we have that
\begin{align}
	\a_0 &= \frac{4 c \conj{\kappa} \conj{\mu}^2}{m^* \left(4 |m \mu|^2 \lambda ^2+1\right) \left(5 c^6+4 \lambda ^2 \left( |\kappa|^2 c^6+\left(c^6+2\right) |m \mu|^2 \right)+2\right)} \nonumber\\
	&\times \Bigg(-32 \lambda ^4 |\mu|^4 |m|^6 +4 \lambda^2 \mu  \left(|\kappa|^2 \lambda ^2 c^6 + c^6-4\right) \conj{\mu} |m|^4 \nonumber\\
	&+ |m|^2 \left(\lambda ^2 \left(4 \lambda ^2 |\kappa|^4 + 9 |\kappa|^2 \right) c^6+5 c^6-2\right) - |a|^2 \left( |\kappa|^2 \lambda ^2+1\right) \left(4 |m \mu|^2 \lambda ^2+1\right)\Bigg) \nonumber\\
	\a_1 &= -\frac{2 \kappa  \mu ^2 \left(m^*\right)^2}{c}.
\end{align}

\section{Further SU(2) T-brane backgrounds}
\label{ap:more}

We have analysed in section \ref{s:Tbrane} two different cases of T-brane backgrounds, whose non-commuting Higgs field generators lie entirely within an $\mathfrak{su} (2)$ subalgebra of the Lie group. As discussed in section \ref{s:gen}, whenever that is the case one may focus on such $\mathfrak{su} (2)$ subalgebra  when solving for the $\a'$-corrected D-term equations, as the equations corresponding to other generators decouple. In this appendix we will apply the analysis of section \ref{s:Tbrane} to further $SU(2)$ T-brane backgrounds, which are also examples of the $2 \times 2$ T-brane blocks discussed in section \ref{s:gen}. Unlike the examples in section \ref{s:Tbrane}, here none of the backgrounds will be associated to a monodromy. In general we find that the presence or absence of monodromy does not really affect the behaviour of $\a'$-corrections in T-brane systems.

In general we will follow the strategy of subsection \ref{s:Tbrane/SU2} when analysing the backgrounds below. First we consider an Ansatz with a gauge transformation of the form (\ref{gf}) with $f \equiv f(x,\xb,y,\yb)$ and a worldvolume flux of the form (\ref{H1}). In general we find that the Ansatz for the gauge transformation can be reduced to $f \equiv f(x,\xb)$. Moreover the effect of $\kappa$ can be absorbed in the parameter $m'$ defined in (\ref{mprime}) in some cases, like in the T-brane examples $1$ and $2$, while others like T-brane example $3$ seem to require a vanishing $\kappa$ or a non-Cartan gauge transformation (c.f. Appendix \ref{ap:genAnsatz}). Second we generalise our flux background to the form (\ref{Hflux}) and consider the expansion (\ref{expf}) for the gauge transformation, which in practice result in functions $f$ and $h$ that only depend on $(x, \xb)$, at least at lowest order in the expansion parameter $\lam \rho$. As the procedure is identical for all the cases we present our results in a sketchy way, displaying the independent D-term equations for each Ansatz and the asymptotic solutions near the origin for the second one. All of the following examples satisfy $[\phi, \phib] \equiv C \sigma _3$ for some $C$ depending on the Higgs-vev, which we will use to abbreviate the following expressions. We will compute the D-term equations for the same two Ans\"atze as in \ref{s:Tbrane}. That is, on the one hand for a flux consisting of the two components
\begin{align}
	F &= -i \p \pb f \cdot \sigma_3  & f\equiv f(x,\xb,y,\yb) \nonumber\\
	H &= \im \left(\kappa \, dx \wedge d\bar{y}\right)\, \mf{1},
\end{align}
henceforth called Ansatz 1, and on the other hand for
\begin{align}
	F &= -i \p \pb f \cdot \sigma_3 \nonumber\\
	H &= \im \left(\kappa \, dx \wedge d\bar{y}\right)\, \mf{1} +\rho \,  i \left(dx \wedge d\bar{x} - dy \wedge d\bar{y} \right) \, \mf{1} - i \p\bar{\p} h \, \mf{1} \nonumber\\
	f &\equiv f(x,\xb) \, h \equiv h(x,\xb,y,\yb),
\end{align}
called Ansatz 2 in the following.

\subsection*{T-brane 1}

\begin{align}
	\phi_\mr{hol} = m\left(\begin{matrix}
		0 & 1 \\ 0 & 0
	\end{matrix}\right)
\end{align}
\paragraph{Ansatz 1:}
\begin{align}
	(\p_x \pb_\xb + \p_y \pb_\yb) f &= C \left( 1 + \lambda^2 |\kappa|^2 + 4 \lambda ^2 Q_f \right) \nonumber\\
	&-\frac{8}{3} \lambda ^2 |m|^2 e^{2 f} \left(\p_y \pb_\yb f  \pb_\xb f  \p_x f -\p_y f  \pb_\xb f  \p_x \pb_\yb f +\pb_\yb f  \left(\p_y f  \p_x \pb_\xb f -\p_y \pb_\xb f  \p_x f \right)\right) \nonumber
\end{align}
\paragraph{Ansatz 2:}
\begin{align}
	\p_x \pb_\xb f &= C \left( 1 + 4\lambda ^2 Q_H \right) \nonumber\\
	(\p_x \pb_\xb + \p_y \pb_\yb) h &= -8 \lambda ^2 |m|^2 e^{2 f} \pb_\xb f  \p_x f  \left(\p_y \pb_\yb h +\rho \right) +4 C \lambda ^2 \p_x \pb_\xb f ( \rho + \p_y \pb_\yb h) \nonumber
\end{align}
\paragraph{Asymptotic solution}
\begin{align}
	f_0 &= \log c + c^2 |m'x|^2 + \frac{1}{2} c^4 |m'x|^4 \nonumber\\
	f_1 &= -4 |mx|^2 \left(4 c^6 \lambda ^2 |m|^2 |m'|^2 +c^2 \right) -4 c^4 |m'|^2|m|^2 x|^4 \left( 10 c^4 \lambda ^2 |m'|^2 |m|^2 +1 \right) \nonumber\\
	h &= -4 c^4 \lambda ^2 \rho |m|^2 |m'x|^2 -6 c^6 \lambda \rho |m|^2 |m'x|^4 \nonumber
\end{align}

\subsection*{T-brane 2}

\begin{align}
	\phi_\mr{hol} = m\left(\begin{matrix}
		0 & ax \\ 0 & 0
	\end{matrix}\right)
\end{align}
\paragraph{Ansatz 1:}
\begin{align}
(\p_x \pb_\xb + \p_y \pb_\yb) f &= C \left( 1 + \lambda^2 |\kappa|^2 + 4 \lambda ^2 Q_f \right) \nonumber\\
&-\frac{2}{3} \lambda^2 |ma|^2 e^{2 f} \Big(\p_y \pb_\yb f \left|2 x \p_x f +1\right|^2 +4 |x|^2 |\p_y f|^2 \p_x \pb_\xb f \nonumber\\
&-4 \re \left( x \p_y f  \left(2 \xb \pb_\xb f +1\right) \p_x \pb_\yb f \right) \Big) \nonumber
\end{align}
\paragraph{Ansatz 2:}
\begin{align}
	\p_x \pb_\xb f &= C \left( 1 + 4\lambda ^2 Q_H \right) \nonumber\\
	(\p_x \pb_\xb + \p_y \pb_\yb) h &= -2 \lambda^2 |ma|^2 e^{2 f} \left|2 x \p_x f +1\right|^2 \left(\p_y \pb_\yb h +\rho \right) +4 \lambda^2 C \p_x \pb_\xb f ( \rho + \p_y \pb_\yb h) \nonumber
\end{align}
\paragraph{Asymptotic solution}
\begin{align}
f_0 &= \log c + \frac{1}{4} c^2 |m'a|^2 |x|^4 \nonumber\\
f_1 &= -c^2 |am|^2 |x|^4 \left(2 \lambda ^2 c^2 |am|^2 +1\right) \nonumber\\
h &=  -2 \lambda^2 \rho c^2 |amx|^2 \nonumber
\end{align}

\subsection*{T-brane 3}
\label{ap:more/3}

\begin{align}
	\phi_\mr{hol} = m\left(\begin{matrix}
		by & ax \\ 0 & by
	\end{matrix}\right)
\end{align}
\paragraph{Ansatz 1:}
\begin{align}
	\left( 1 + 4 \lambda^2 |mb|^2 \right) \p_x \pb_\xb f -\p_y \pb_\yb f &=  C \left( 1 + \lambda^2 |\kappa|^2 + 4 \lambda ^2 Q_f \right) \nonumber\\
	&-\frac{2}{3} \lambda^2 |ma|^2 e^{2 f} \Big(\p_y \pb_\yb f \left|2 x \p_x f +1\right|^2 +4 |x|^2 |\p_y f|^2 \p_x \pb_\xb f \nonumber\\
	& -2 \re \left( x \p_y f  \left(2 \xb \pb_\xb f +1\right) \p_x \pb_\yb f \right) \Big) \nonumber\\
	0 &= -i \lambda^2 a\conj{b} |m|^2 \conj{\kappa} e^{f} \left(2 x \p_x f +1\right) \nonumber
\end{align}
\paragraph{Ansatz 2:}
\begin{align}
	\p_x \pb_\xb f  \left(1 + 4 \lambda^2 |mb|^2 \right) &= C \left( 1 + 4\lambda ^2 Q_H \right) \nonumber\\
(\p_x \pb_\xb + \p_y \pb_\yb) h &= -2 \lambda ^2 |m|^2 \left(|a|^2 e^{2 f} \left|2 x \p_x f +1\right| \left(\p_y \pb_\yb h +\rho \right)+2 |b|^2 \left(\p_x \pb_\xb h -\rho \right)\right) \nonumber\\
& +4 \lambda^2 C \p_x \pb_\xb f ( \rho + \p_y \pb_\yb h) \nonumber\\
	0 &= \left(2 x \p_x f +1\right) \left(2 \p_y \pb_\xb h +\conj{\kappa}\right) \nonumber
\end{align}
\paragraph{Asymptotic solution}
\begin{align}
f_0 &= \log c + \frac{|am|^2 |x|^4 c^2}{16 \lambda ^2 |bm|^2 +4} \nonumber\\
f_1 &= -c^2 |am|^2 |x|^2 \left(2 \lambda ^2 c^2 |ma|^2 +1\right) \nonumber\\
h &= -\frac{2 \lambda^2 \rho |mx|^2 \left( c^2 |a|^2 -2 |b|^2 \right)}{4 \lambda ^2 |bm|^2 +1} \nonumber
\end{align}

\subsection*{T-brane 4}
\begin{align}
\phi_\mr{hol} &= m\left(\begin{matrix}
0 & ax \\ by & 0
\end{matrix}\right)
\end{align}
\paragraph{Ansatz 1:}
\begin{align}
	(\p_x \pb_\xb + \p_y \pb_\yb) f &= C \left( 1 + \lambda^2 |\kappa|^2 + 4 \lambda ^2 Q_f \right) \nonumber\\
	&-\frac{2}{3} \lambda^2 |ma|^2 e^{2 f} \Bigg(\p_y \pb_\yb f \left|2 x \p_x f +1\right|^2 +4 |x|^2 |\p_y f|^2 \p_x \pb_\xb f  \nonumber\\
	& -4 \re \left(x \p_y f \left(2 \xb \pb_\xb f +1\right) \p_x \pb_\yb f \right) \Bigg) \nonumber\\
	&-\frac{2}{3} \lambda^2 |mb|^2 e^{-2 f} \Bigg( \p_x \pb_\xb f \left| 2 y \p_y f - 1 \right|^2 + 4 |y|^2 |\p_x f|^2 \p_y \pb_\yb f  \nonumber\\
	& +4 \re \left( y \left(1-2 \yb \pb_\yb f \right) \p_y \pb_\xb f  \p_x f\right) \Bigg) \nonumber\\
	0 &= |a|^2 e^{2 f} \re \left( \kappa  x \p_y f  \left(2 \xb \pb_\xb f +1\right)\right) +|b|^2 e^{-2 f} \re \left( \kappa  \yb \pb_\xb f \left(2 y \p_y f -1\right) \right) \nonumber
\end{align}

\end{document}